\documentclass[11pt]{article}
\usepackage{amssymb,amsmath, amsthm}
\usepackage{dsfont}
\usepackage[mathscr]{euscript}
\usepackage{url}
\usepackage{setspace} 
\usepackage{lscape} 
\usepackage{subfig} 
\usepackage{multirow}
\usepackage{lscape}
\usepackage{enumerate}
\usepackage{float}
\usepackage{graphicx}
\usepackage{epstopdf}
\usepackage{centernot}
\usepackage[labelfont={bf}]{caption} 
\usepackage{adjustbox}
\usepackage{tikz}\usetikzlibrary{arrows,matrix,positioning,fit}
\usepackage{fullpage}
\usepackage{marginnote}

\makeatletter
\let\@fnsymbol\@arabic
\makeatother 

\title{Dynamic communicability and epidemic spread: a case study on an empirical dynamic contact network}

\author{Isabel Chen\thanks{Department of Mathematics and Computer Science, Emory University, Atlanta, Georgia 30322, USA (imchen@mathcs.emory.edu).} \and Michele Benzi\thanks{Department of Mathematics and Computer Science, Emory University, Atlanta, Georgia 30322, USA (benzi@mathcs.emory.edu).  The work of this author was supported by National Science Foundation grant DMS-1418889.} \and Howard H. Chang\thanks{Department of Biostatistics and Bioinformatics, Rollins School of Public Health, Emory University, Atlanta, Georgia 30322, USA (howard.chang@emory.edu).} \and Vicki S. Hertzberg\thanks{Center for Nursing Data Science, Nell Hodgson Woodruff School of Nursing, Emory University, Atlanta, Georgia 30322, USA (vhertzb@emory.edu).}}

\date{}

\begin{document}
\maketitle

\begin{abstract}

We analyze a recently proposed temporal centrality measure applied to an empirical network based on person-to-person contacts in an emergency department of a busy urban hospital. We show that temporal centrality identifies a distinct set of top-spreaders than centrality based on the time-aggregated binarized contact matrix, so that taken together, the accuracy of capturing top-spreaders improves significantly. However, with respect to predicting epidemic outcome, the temporal measure does not necessarily outperform less complex measures. Our results also show that other temporal markers such as duration observed and the time of first appearance in the the network can be used in a simple predictive model to generate predictions that capture the trend of the observed data remarkably well.
\end{abstract}
\section{Introduction}
The presence of infectious agents in a confined space bring substantial risks of cross infection. In this paper we make use of network analysis to better understand contagion processes within such spaces. We base our study on the interactions of people in an Emergency Department (ED) of a hospital in the Midtown area of Atlanta, GA \cite{VH-PLOS}.   

Utilizing network structure in epidemiological models relaxes the assumption that interactions between agents are well-mixed \cite{spatialEpi, networks+epi3, spatialEpi2, networks+epi2}. The study of empirical contact data has also revealed patterns that differ from a priori contact assumptions \cite{mossong, readJM}, which highlights the need to better understand real-world data. Recent technological advances have made high-dimensional data available for study, and thorough analyses of empirical temporal network data can be found in \cite{cattuto, VH-PLOS, readJM, Rocha-SexNetwork} and \cite{Stehle}. 

\begin{figure}[H]
\centering
\caption{An example of a dynamic network}
\includegraphics[scale=0.7]{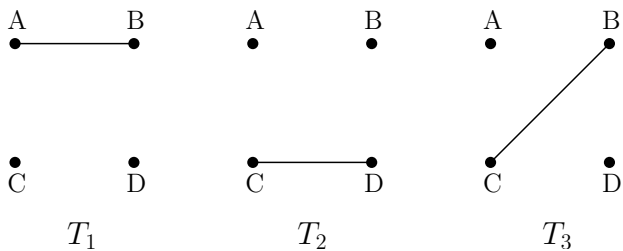}
\label{dynamic_network}
\end{figure}

In the context of disease spread, temporal information plays a crucial role \cite{bansal}. The length of contact time between agents has been shown to be important in the contagion process \cite{readJM, smieszek, Toth}. The temporal order of contact sequences is also crucial: a person can only pass on the disease to others after becoming infectious. For illustration, consider the following time-evolving network on four nodes shown in \ref{dynamic_network}, with node A being the only infectious node at time $T_1$. An edge between two nodes is representative of some form of disease-transmitting contact between them. 

\begin{figure}[H]
\centering
\caption{The aggregated version of Figure \ref{dynamic_network}}
\includegraphics[scale=0.7]{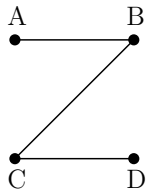}
\label{agg_network}
\end{figure}

For simplicity, we assume that an infected person is immediately infectious. There is a temporal path $A \rightarrow B \rightarrow C$ which means that the disease can potentially spread from $A$ to both $B$ and $C$. On the other hand, it is impossible for $D$ to become infected, because the contact between $C$ and $D$ occurs \emph{before} $C$ has contact with $B$. Note that the static version of the network (Fig \ref{agg_network}) loses such information: without the temporal dimension, all nodes, including $D$, are reachable from $A$. Work in \cite{Fefferman} and \cite{Volz+Meyers} have shown that the non-constancy of contact structure can have a significant impact on disease spread, and should therefore be incorporated into the disease model if such information is available.     


An important objective is to identify key players in the infection process, and centrality measures are a natural choice to consider for this task. The relationship between centrality and epidemic outcome has been studied in \cite{role_of_centrality, bauer_lizier, kitsak, Salathe} and \cite{renato}, but the networks under study did not include temporal information. It has also been argued that while some centrality measures are able to identify highly influential nodes, they do not accurately quantify \cite{bauer_lizier,influence_of_all_nodes}, and may even underestimate \cite{centrality_underestimates} the spreading power of the vast majority of nodes which are not highly influential. Alternative spreading power metrics such as the accessibility \cite{accessibility2,accessibility} and expected force \cite{influence_of_all_nodes}, which extend centrality by incorporating spreading dynamics, have been shown to have stronger correlation with epidemic outcome than traditional centrality \cite{bauer_lizier,
klemm_serrano_egulutz_miguel, renato}. Before rejecting centrality measures completely as a means to understand contagion on networks, we seek to answer the question: will incorporating temporal information into centrality metrics improve their explanatory and/or predictive power in relation to epidemic outcome? 

We are motivated by examples in \cite{DynamicCommunicators} and \cite{Tang} which show that centrality measures based on time-aggregated versions of temporal networks (as shown in Fig. \ref{agg_network}) fail to adequately capture important nodes. The work in \cite{Rocha-SexNetwork} suggests that temporal correlations in network data should not be underestimated and that consequently, detecting important individuals based on temporal structure may have a significant impact on targeted intervention strategies. In \cite{Salathe} it was shown that while centrality measures on the time-aggregated network (such as node degree, betweenness centrality, eigenvector centrality) can improve immunization strategies, node strength, which measures the total time exposed to others, was the most effective.  These results illustrate the importance of utilizing temporal information in the context of identifying key players in the contagion process. 

Our work aims to bring recently developed temporal centrality measures into the analysis. Temporal centrality measures provide a means of identifying important nodes based solely on time-dependent network structure. While temporal centrality is a relatively young field compared to its static counterpart, new measures are constantly being developed \cite{modern_temporal_network_theory, temporal_node_centrality, Tang, eig_based_temporal_centrality}. We focus our attention on \emph{dynamic communicability} introduced by Grindrod \emph{et al.} \cite{Grindrod}. This is a generalization of Katz centrality \cite{Katz} to the dynamic setting. We emphasize that the computation of centrality scores depend solely on the network structure and is completely independent of any contagion process on the network. 




We simulate an epidemic process on the network, where the duration of contacts is used explicitly in the infection procedure. We aim to address the following two questions: First, what role does temporal centrality play in explaining epidemic outcome? Second, can temporal centrality be used to predict epidemic outcome? It has often been asserted in the literature \cite{role_of_centrality,kitsak, klemm_serrano_egulutz_miguel, influence_of_all_nodes, renato} that a strong correlation between epidemic outcome ($Y$) and some measure, say $X$, will result in measure $X$ being a good predictor for $Y$. The notion of correlation measures the strength of linear relationship between $X$ and $Y$, and is intrinsically a goodness-of-fit measure of the dataset under consideration. Determination of predictive power, on the other hand, must be assessed on a \emph{different} dataset. While a strong correlation is suggestive of good predictive power, correlation will not reflect the predictive power between variables with a non-linear relationship. In our work, instead of looking at raw correlations, we use multiple linear regression to study the overall effect of centrality on epidemic outcome. We recognize that centrality measures alone cannot fully explain epidemic outcome, and the regression framework allows for the inclusion of other variables into the analysis in order to gain a better understanding of the role that dynamic network centrality plays in the epidemic process. The regression model generates coefficients that quantify the relationship between predictor and response, and it is these coefficients that are used in the prediction process, not the correlation itself. Predictions are performed on different samples of networks drawn from the same study. This provides a more accurate assessment of the predictive power of centrality in the context of epidemic spread. 

Employing network centrality to explain node characteristics/behavior has been done in the social sciences \cite{christakis_depression, christakis_latrine}, where path-based measures, such as betweenness and closeness centrality, are typically used. Our goal in this work is two-fold: first, to examine the effectiveness of \emph{temporal network centrality} in explaining and/or predicting epidemic outcome, and second, to highlight temporal, \emph{walk-based} measures such as dynamic communicability, which, due to their formulation in terms of matrix functions, can be computed with greater computational ease than path-based measures.

The rest of the paper is organized as follows: In Section \ref{network} we describe the temporal network under consideration. In Section \ref{DC} we introduce the notion of dynamic communicability and in Section \ref{DC-results} we present the results of dynamic communicability applied to our dataset. Section \ref{InfModel} describes the infection model used to simulate disease spread on the temporal network, and in Section \ref{SimResults} we present the simulation results. In Section \ref{section: virulence+centrality} we analyze the relationship between virulence and centrality. We conclude and discuss future directions in Section \ref{Conclusion}.

\section{Description of temporal network}\label{network}
The Emergency Department (ED) of Emory University Hospital (Midtown, Atlanta) was divided into 95 zones. The zones were designed in such a way that two people in the same zone are within 1 meter of each other with very high probability. Although no physical contact is guaranteed, the close proximity of any two people within a zone is considered a potential disease-spreading contact. Such close proximity interactions are particularly pertinent to infections that are transmitted predominantly via droplets. Participants in the study wear radio-frequency (RFID) tags which track their movements in the ED. Both patients and staff were recruited to participate. Three groups of staff were present: medical doctors (MD), registered nurses (RN) and administrative staff (other). Data was collected over 35 shifts of at most 12 hours. See \cite{VH-PLOS} for more details on the data collection procedure.

A temporal network is constructed based on the movements of the participants in the ED. The nodes represent people and the edges represent their interactions within zones. The dynamic network is viewed as a sequence of adjacency matrices on the same set of nodes, where each matrix represents the connections present at the corresponding time-step. The time-evolution of the network is thus captured by the appearance and disappearance of edges over time. 

Two people who are in the same zone during a 1-second time-frame are said to share a location. Two people who are in the same zone during a 1-second time-frame are said to share a location. The RFID tags transmitted their unique identifier every 10 seconds (\cite{VH-PLOS}). For this reason, the per-second data is considered incomplete and therefore we work with adjacency matrices at the 10-second time resolution. Explicitly, the full 12-hour shift is divided into 10-second intervals, labeled $t=1,\ldots,4321$. For each value of $t$ we have an adjacency matrix which aggregates the contact information within this 10-second time interval:
\begin{align*}
A^{[t]}_{ij} &:= \left\{\begin{array}{cl}
1 & \mbox{if person $i$ and person $j$ shared a location within the 10-sec interval $t$} \\
0 & \mbox{otherwise}
\end{array}
 \right. 
\end{align*}
We point out that the total contact time between two people within the 10-second interval is not taken into account. It is possible that two people were in the same location for longer than 1 second, or that they crossed paths at different locations within the 10-second interval.

We analyze the contact data obtained over 7 different shifts. Shift-specific information is shown in Table \ref{table: shift_data}. 

\begin{table}[h!]
\centering
\footnotesize
\arraycolsep=1.4pt\def\arraystretch{1.4}
\caption[Shift-specific data]{Shift-specific data. Analysis of dynamic communicability and its relationship to epidemic outcome is performed on Shift 1 (training data). Predictive power is assessed on Shifts 2 to 7 (testing data). Participation rate (p. rate) is the percentage of people present in the ED who participated in the study.}
\label{table: shift_data}
\begin{tabular}{ccccccccc}\hline\hline																	
Shift	&	$n$	&	staff	&	patients	&	total patients	&	participation rate ($\%$)	&	shift length	&	am/pm	&	weekday	\\\hline
1	&	107	&	33	&	74	&	98	&	76	&	11	&	pm	&	y	\\
2	&	115	&	33	&	82	&	117	&	70	&	12	&	pm	&	y	\\
3	&	89	&	25	&	64	&	82	&	78	&	8	&	am	&	n	\\
4	&	129	&	34	&	95	&	108	&	88	&	12	&	pm	&	n	\\
5	&	133	&	44	&	89	&	117	&	76	&	11.75	&	pm	&	y	\\
6	&	87	&	26	&	61	&	77	&	79	&	8	&	am	&	n	\\
7	&	126	&	35	&	91	&	133	&	68	&	11.67	&	am	&	n	\\ \hline \hline
\end{tabular}

\end{table}

\section{Dynamic communicability}
\label{DC}
Dynamic communicability \cite{Grindrod} is a network centrality measure that defines a node's importance based on the dynamic walks it participates in. A dynamic walk\footnote{We emphasize that the same edge can be used multiple times in a \emph{walk}, in contrast to a \emph{path}, where edges must be distinct. Betweenness centrality and closeness centrality are path-based measures, and temporal versions are based on the analogous notion of \emph{temporal paths} -- see for example \cite{temporal_node_centrality}. Path-based metrics are typically computationally more intensive than walk-based measures.} is a sequence of edges connecting nodes, with the added constraint that the sequence of edges must respect the time ordering. The rationale is that nodes which participate in many dynamic walks are capable of communicating well with other nodes in the network, and are therefore potential candidates for effective spreading or collecting of information, where information is referred to in the broadest sense possible. In addition, within this framework, short walks are given more importance than long walks, since fewer edges allow for faster and less noisy transmission of information. 

Consider a time-evolving network on $n$ nodes, which is represented by a sequence of adjacency matrices $A^{[k]}$ for $k=1,\ldots, M$, where $k$ indexes the time-step. For each adjacency matrix, the spectral radius is defined as $\rho\left(A^{[k]}\right) = \max\{|\lambda_1|,\ldots,|\lambda_n|\}$ where the $\lambda$'s are the eigenvalues associated with $A^{[k]}$. Let $\sigma^* = \max_k\left\{\rho\left(A^{[k]}\right)\right\}$ be the maximum spectral radius over all the adjacency matrices. The dynamic communicability matrix is defined as the product of matrix resolvents 
\begin{equation}
\label{Eq: Q}
Q = \left(I-\alpha A^{[1]} \right)^{-1} \left(I-\alpha A^{[2]} \right)^{-1} \cdots \left(I-\alpha A^{[M]} \right)^{-1} 
\end{equation}
where, in order to ensure that each resolvent can be expressed as a power series in the matrix, the parameter $\alpha$ must satisfy $0<\alpha<1/\sigma^*$. Expressing each resolvent as a power series and expanding out the product, we see that $Q$ contains all products of the form 
\begin{equation}
\label{dynamic walks}
\alpha^w \left( A^{[t_1]}A^{[t_2]}\cdots A^{[t_w]} \right),
\end{equation}
where $t_1\leq t_2\leq\cdots\leq t_w$. Observe that the $ij$th entry of (\ref{dynamic walks}) counts the number of dynamic walks of length $w$ from node $i$ to node $j$, where the $k$th edge of the walk comes from time-step $t_k$.  Furthermore, this count is downweighted by $\alpha^w$. The restriction that $t_1\leq t_2\leq\cdots\leq t_w$ ensures that the dynamic walks are legitimate in the physical sense: subsequent edges in the walk cannot come from an earlier time-step. Consequently, $Q_{ij}$ is a weighted sum of dynamic walks of all possible lengths between nodes $i$ and $j$, where walks of length $w$ are downweighted by $\alpha^w$. This is a measure of `communicability' between node $i$ and node $j$, with node $i$ being the broadcaster and node $j$ being the receiver. Let $\mathds{1}$ denote the vector of all ones. Summing over all receivers $j$, we obtain 
\[\sum_{j=1}^n Q_{ij}= \bigg( Q\cdot \mathds{1}\bigg)_i\]
which is a measure of how well node $i$ broadcasts information to the rest of the network as a whole. The row sums therefore provide a notion of \emph{broadcast centrality} (BC). On the other hand, the column sums
\[\sum_{i=1}^n Q_{ij}= \bigg( Q^T\cdot \mathds{1}\bigg)_j\]
measures how well node $j$ receives information from all other nodes in the network, and therefore provides a notion of \emph{receive centrality} (RC). The matrix $Q$ is therefore able to capture dual notions of broadcasting and receiving. We reiterate that $Q$ is in general not symmetric, therefore the row and column sums are in general different from each other. We also point out that because the base matrices $A^{[k]}$ are symmetric, the resolvents $(I-\alpha A^{[k]})^{-1}$ are also symmetric. Furthermore, since for any square matrix $A$, $\left(A^{-1}\right)^T = \left(A^T\right)^{-1}$, we have 
\begin{align*}
Q^T & = \bigg( \left(I-\alpha A^{[1]} \right)^{-1} \left(I-\alpha A^{[2]} \right)^{-1} \cdots \left(I-\alpha A^{[M]} \right)^{-1} \bigg)^T \\
& = \left(I-\alpha A^{[M]} \right)^{-1} \left(I-\alpha A^{[M-1]} \right)^{-1} \cdots \left(I-\alpha A^{[1]} \right)^{-1}.
\end{align*}
In other words, broadcast and receive centralities are related by a reversal of the time ordering. 

Synthetic examples in \cite{Grindrod} and \cite{DynamicCommunicators} illustrate that broadcast centrality (BC) and receive centrality (RC) measures perform better than aggregated measures in identifying nodes with time-sensitive links as important. The term \emph{dynamic communicator} coined in \cite{DynamicCommunicators} refers precisely to the nodes which rank highly in the dynamic sense but do not stand out in a snapshot or aggregate view of the network. 

\subsection{Limit as $\alpha \rightarrow 0$} \label{section:Limit}
Expressing each resolvent as a power series, then expanding the product and collecting terms, we see that 
\[Q = \left(I  -\alpha A^{[1]} \right)^{-1} \ldots \left(I-\alpha A^{[M]} \right)^{-1} = I + \alpha \left( \sum_{k=1}^M A^{[k]}\right) + \mathcal{O}(\alpha^2).\] 
Letting $BC$ denote the vector of broadcast centrality measures, we have
\[BC := Q\cdot \mathds{1} = I\cdot \mathds{1} + \alpha \left( \sum_{k=1}^M A^{[k]}\right)\cdot \mathds{1} + \mathcal{O}(\alpha^2).\]
Shifting by $\mathds{1}$ and scaling by the constant $\alpha$, we have 
\begin{align*}
 \frac{BC- \mathds{1}}{\alpha} = \frac{Q \cdot \mathds{1} - \mathds{1}}{\alpha} = \underbrace{\left( \sum_{k=1}^M A^{[k]}\right) \cdot \mathds{1}}_{AD} + \mathcal{O}(\alpha)
\end{align*}
Observe that the first term on the RHS is a measure of aggregated degree, denoted by AD, which ranks nodes according to the number of distinct contacts weighted by the duration of contact time, in the sense that the longer the contact time between a pair of nodes, the higher their corresponding AD values. (This is in contrast to binarized degree BD, where no temporal information is captured -- see Section \ref{section: virulence+centrality}.) Since a shift and scale of the BC vector will not change the associated BC rankings, we see that as $\alpha \rightarrow 0$, BC rankings should theoretically converge to AD rankings. A similar argument shows that the same is true for RC rankings. We will use this fact as a test for the numerical accuracy of our results, as well as to inform our choice for the parameter $\alpha$. Recall that $\alpha$ must be chosen so that 
$0<\alpha< 1/\sigma^*$. We aim to choose $\alpha$ sufficiently far away from $0$ so that we do not merely replicate aggregate degree. On the other hand $\alpha$ cannot be too close to the upper limit, as in this case, the matrix for which the maximum spectral radius is attained will be close to singular, and the entries of its inverse will dominate the computation of $Q$. In this regime, the computation of $Q$ will be sensitive to small changes in $\alpha$. See also \cite{Benzi+Klymko} for further discussion on the role of $\alpha$ as a tuning parameter in the context of non-temporal networks.  

\subsection{Possible modifications} 

As discussed in \cite{Grindrod}, in order to eliminate the possibility of long walks, or closed walks such as $i \mapsto j \mapsto i$ taking place within a single time-step, we may enforce the walks to use at most one link per time window by using this modified version of $Q$ where each term in the product is a first order approximation of the matrix resolvent:
\[\hat{Q} = (I+\alpha A^{[1]})(I+\alpha A^{[2]})\cdots (I+\alpha A^{[M]}).\] 
This contains all products of the form $\alpha^w A^{[t_1]}\ldots A^{[t_w]}$ where $t_1<\cdots<t_w$ are all distinct. This modification may be appropriate in ultra-high frequency regimes, where the length of each time window is so small that it is physically unfeasible for information to pass through more than one link per time step.  

\subsection{Data studied} 

Dynamic communicability has been studied on telecommunication data (MIT \cite{Grindrod}) and
email data Enron \cite{Grindrod}, \cite{DynamicCommunicators}. It has also been used to characterize learning in the human brain \cite{Brain}. We would like to study BC/RC measures in the context of disease spread on a person-to-person contact network. In particular, we want to see how a seed node's BC ranking/measure is related to epidemic outcome. We point out that \cite{Higham_predicts} studies the same measure, dynamic communicability, in relation to contagion on a temporal network. Our work differs in the methodology of the infection simulation, as well as in the use of regression to analyze the resulting simulation output. Additionally, the network studied in \cite{Higham_predicts} is based on email communication, on which the notion of epidemic spread is inherently different from that of a proximity-based contact network.       
In \cite{CompleNet}, dynamic communicability was used to find a subset of nodes to maximize influence on the SocioPatterns hospital ward dynamic contact network. Our work does not focus only on the highly central nodes, but aims to evaluate the overall effect of dynamic communicability on epidemic outcome. 

\subsection{Relationship with the matrix exponential}

One can also use the matrix exponential instead of the matrix resolvent to compute a measure of dynamic communicability \cite{matrixExpDC}. The rationale of downweighting long walks remains, with the difference lying in the downweighting factors themselves: $1/w!$ versus $\alpha^w$. One can replace the matrix resolvents with matrix exponentials to obtain another walk-based measure in the following way:
\begin{equation}
\label{Eq: Qexp}
\tilde{Q} = e^{A^{[1]}} e^{A^{[2]}}\cdots e^{A^{[M]}}.
\end{equation}

One difficulty that arises with using the matrix exponential is that the downweighting factor of $1/w!$ typically penalizes walks of length $w$ much more severely than $\alpha^w$. Especially when, in the dynamic case, there are many adjacency matrices to work with, the severe downweighting of long walks may lead to an inability to distinguish between nodes. Preliminary experiments on our data set are indicative of this: Normalization during the computation of $\tilde{Q}$ (see Section \ref{section: computational note}) results in the majority of nodes with broadcast centrality close to zero, and only 2 nodes with non-zero values (see Section \ref{Section: Exp}). On the other hand, dynamic communicability based on the matrix resolvent was able to clearly distinguish about 20 nodes which had broadcast centrality significantly different from zero. 

A possible remedy is to include a tuning parameter $\beta$ in the matrix exponential as follows:
\[e^{\beta A} = I + \beta A + \frac{\beta^2}{2!}A^2 + \ldots \]
The parameter $\beta$ can be viewed, from a thermodynamical point of view, as a form of inverse temperature of the system \cite{matrixExpDC}. Since $\beta$ has no upper bound (unlike $\alpha$ in dynamic communicability), the factor $\beta^k/k!$ can potentially give longer walks more weight than $\alpha^k$. It may be of interest to perform a comparative study of the results obtained by both methods. We will leave this for future work: in the rest of this paper, we discuss results based on the resolvent-based formulation defined in \cite{Grindrod}.

\section{Dynamic communicability applied to the temporal network}
\label{DC-results}
We apply dynamic communicability to the contact data of shift 1 (see Table \ref{table: shift_data}). A total of 107 participants agreed to take part in the study, out of which 33 were staff and 74 were patients. An additional 24 patients were in the ED during that time but did not participate in the study. 

\subsection{Computational note}
\label{section: computational note}
Dynamic communicability as defined in Eq. (\ref{Eq: Q}) can be computed in two ways:
\begin{itemize}
\item Method I: Compute $Q$ explicitly, for example, as suggested in \cite{Grindrod}, using an iteration of the form 
\[\hat{Q}^{[k]} = \frac{\hat{Q}^{[k-1]} \left( I-\alpha A^{[k]}\right)^{-1}}{\Vert \hat{Q}^{[k-1]} \left( I-\alpha A^{[k]}\right)^{-1}\Vert}, \hspace{3cm} k=1,2,\ldots,M \] 
where $\hat{Q}^{[0]}$ is the identity matrix, then compute the row and column sums to obtain BC and RC measures. 
\item Method II: Compute BC and RC measures directly using an iteration of the form 
\[BC^{[k]} = \frac{ \left( I-\alpha A^{[M+1-k]}\right)^{-1}\cdot BC^{[k-1]}}{\Vert \left( I-\alpha A^{[M+1-k]}\right)^{-1}\cdot BC^{[k-1]}\Vert} , \hspace{3cm} k=1,2,\ldots,M \] 
where $BC^{[0]} = \mathds{1}$. A similar form (using the transpose of the resolvents) is used to compute RC measures.  
\end{itemize}
In both methods, normalization is performed to avoid under or overflow in the computations. Here we use the Euclidean $2$-norm, although any matrix norm is applicable. Note that while normalization changes the absolute values of the centrality measures, it does not change the overall rankings of the nodes. For the data based on shift 1, $\alpha_{\max} = 0.072$. 

\begin{table}[h!]
\centering
\arraycolsep=1.4pt\def\arraystretch{1.4}
\caption{Computation times for Method I and Method II}
\begin{tabular}{c|c|c}\hline\hline
choice of $\alpha$	&	Method I (in sec)	&	Method II	(in sec) \\ \hline 
$0.25*\alpha_{\max}$	&	$8.20$	&	$0.70 ~ (\times 2 = 1.40 )$	\\
$0.50*\alpha_{\max}$	&	$17.88$	&	$0.79 ~ (\times 2 = 1.58)$	\\
$0.75*\alpha_{\max}$	&	$9.83$	&	$0.71 ~ (\times 2 = 1.42)$	\\
$0.85*\alpha_{\max}$	&	$9.54$	&	$0.74 ~ (\times 2 = 1.48)$	\\ \hline \hline
\end{tabular}
\label{computationTime}
\end{table}


Computations were done using Matlab. 
For Method I we use \texttt{mrdivide} which solves systems of linear equations of the form $\mathbf{x}A = B$; for Method II we use \texttt{backslash} which solves a linear system of the form $A\mathbf{x} =\mathbf{b}$.  
Table \ref{computationTime} shows that Method II is an order of magnitude faster than Method I. However, because the first iteration of Method II uses the resolvent of the \emph{last} adjacency matrix $A^{[m]}$, in applications where one needs to compute the BC and RC rankings dynamically, Method I is more appropriate. Both methods yield quantitatively similar results. Since, in our application, we do not study how BC/RC rankings themselves change over time, in the following discussion, we will present the results based on the faster method, Method II.   

It is interesting to note that the computation times depend on $\alpha$. We do not have an explanation for this behavior.




\subsection{Robustness with respect to the choice of $\alpha$}
\begin{figure}[h!]
\centering
\caption[Evolution of maximum degree, spectral radius and average degree over time (shift 1)]{Evolution of maximum degree, spectral radius and average degree over time (shift 1, color online). Note that average degree is computed by dividing the total degree by the number of people present in the ED at that time step, and this count includes people who are not necessarily interacting with others.}
\label{fig:plot_spectral_radius}
\includegraphics[width = 0.5\textwidth]{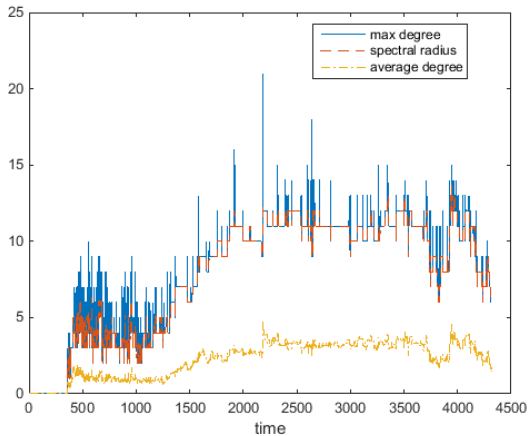}
\end{figure}

Recall that the computation of dynamic communicability requires a choice of the parameter $\alpha$, where 
\[ 0 < \alpha < \frac{1}{\max\limits_k \rho(A^{[k]})} = \alpha_{\max}. \]
For the data based on shift 1, $\alpha_{\max} = 0.072$. In other words, $\max\limits_k \rho(A^{[k]}) \approx 14$ is a lower bound for the maximum degree over all time steps, as illustrated in Figure \ref{fig:plot_spectral_radius}. In this section we present the results obtained for the following choices of $\alpha$: 
\begin{eqnarray*}
\alpha_1 = 0.25 * \alpha_{\max} \\
\alpha_2 = 0.50 * \alpha_{\max} \\
\alpha_3 = 0.75 * \alpha_{\max} \\
\alpha_4 = 0.85 * \alpha_{\max} 
\end{eqnarray*} 

\subsubsection{BC and RC measures}

For interpretative ease, we point out that nodes labeled 1-33 are staff, and nodes labeled 34-107 are patients. We emphasize that the role of centrality metrics is first and foremost to provide a means to rank nodes relative to each other; the numerical values themselves may not be directly interpretable.

\begin{figure}[h!]
\centering
\caption[BC and RC measures (shift 1)]{BC and RC measures (shift 1) associated with different values of $\alpha$. There is one data point per node; the horizontal axis is the node ID label. Nodes labeled 1-33 are staff and nodes labeled 34-107 are patients.}
\subfloat[BC]		
{\includegraphics[width = .5\textwidth]{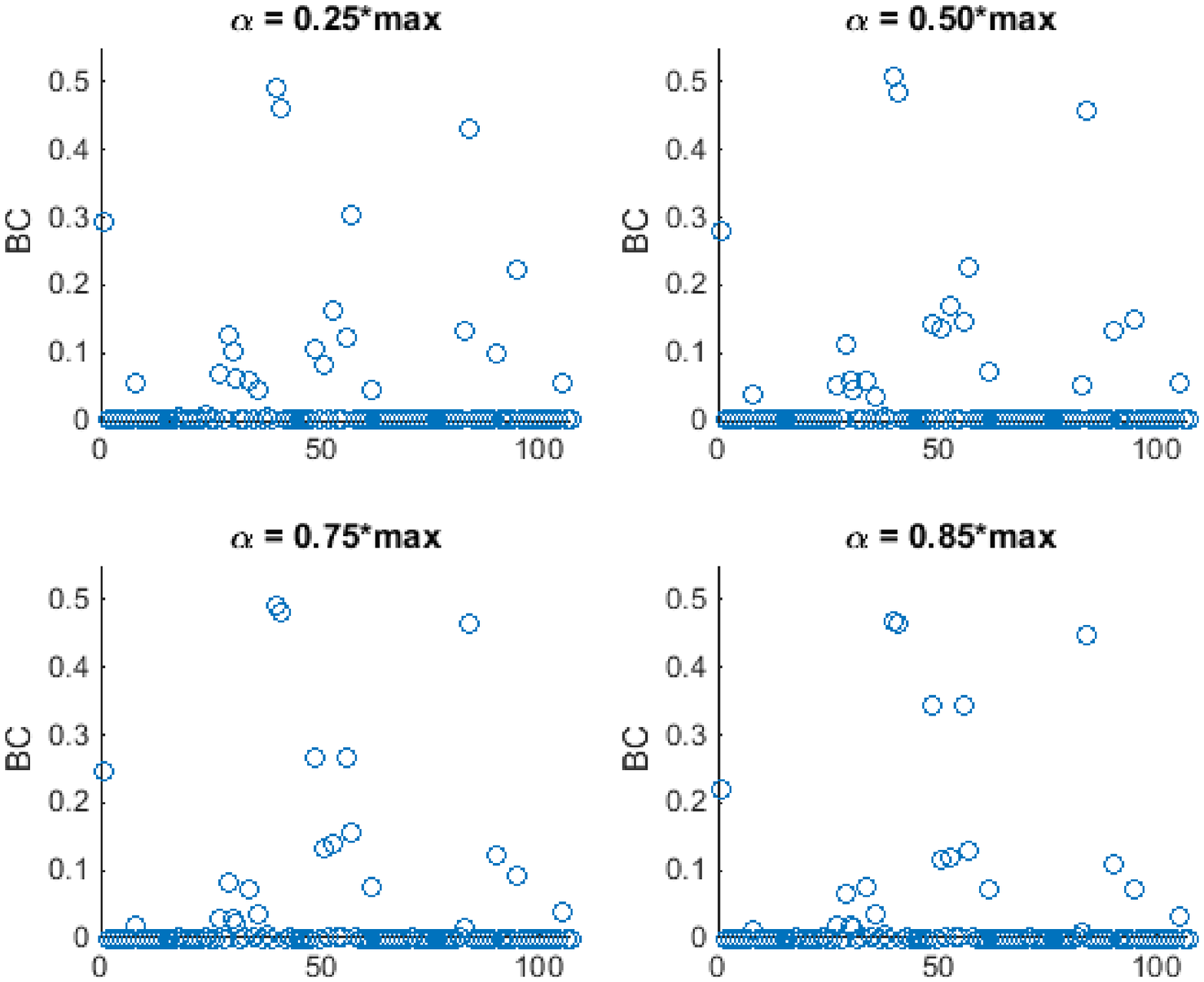}
\label{fig:plotBC}}
\subfloat[RC]
{\includegraphics[width = .5\textwidth]{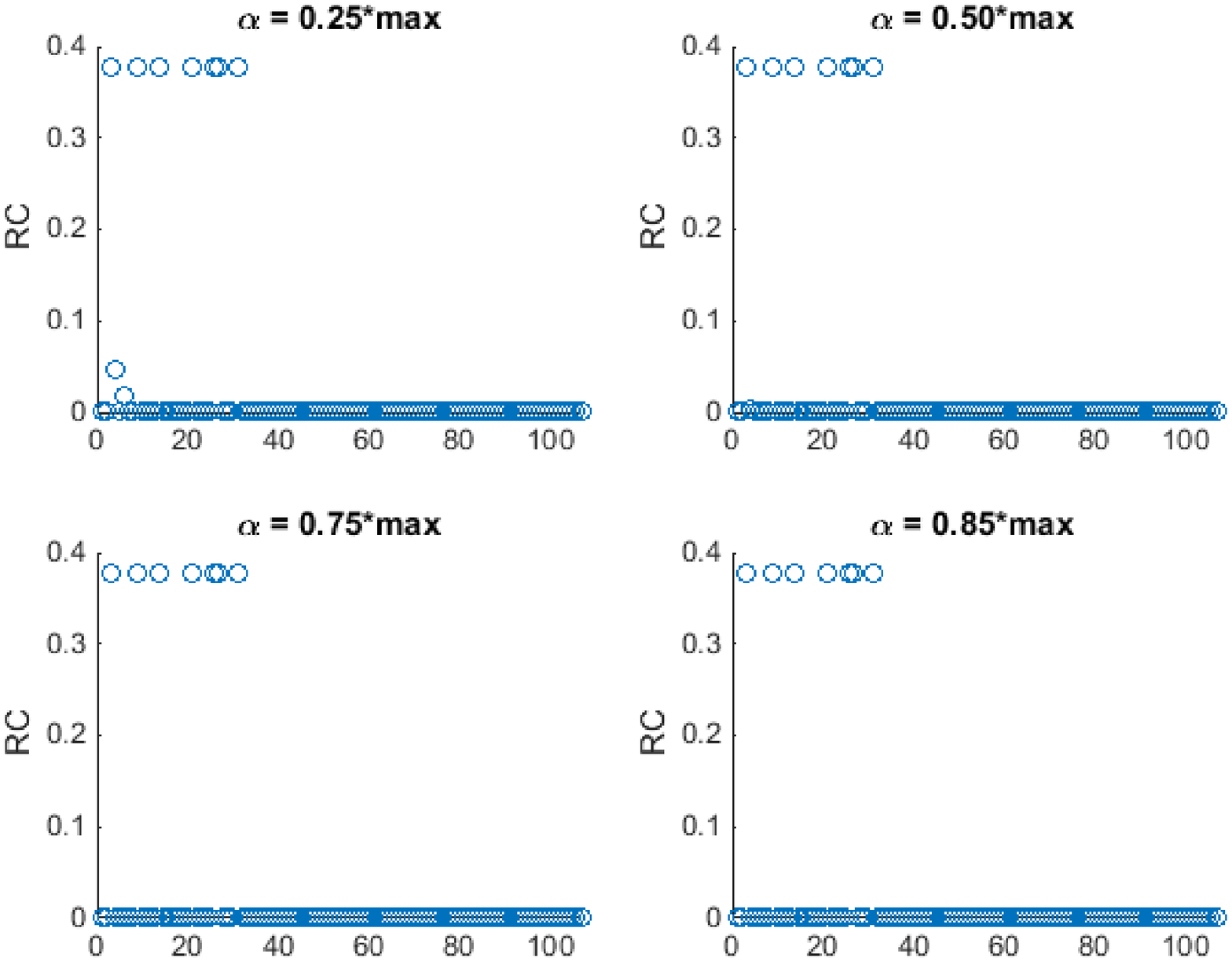}
\label{fig:plotRC}} 
\label{fig:plotBC+RC}
\end{figure}


From Figure \ref{fig:plotBC+RC} we see that most BC and RC measures are close to zero, irrespective of $\alpha$. Although BC and RC measures cannot be exactly zero (if so, $Q$ must have an entire row or column of zeros which is impossible since $Q$ is an invertible matrix), normalization at each iteration in the computation of $Q$ is likely to result in very small values. 

The RC measures exhibit a curious feature: regardless of $\alpha$, the same seven nodes have non-negligible RC measure, and they all have the same magnitude, agreeing to many decimal places. We point out that these nodes are registered nurses (RN). Since RN's are typically the last people that patients see before leaving the ED, it is conceivable that RN's are often at the receiving ends of walks, which explains why the method ranks them as high receivers. 

\begin{figure}[h!]
\centering
\caption[RC measures (shift 1) under different constraints]{RC measures (shift 1) under different constraints. There is one data point per node; the horizontal axis is the node ID label. Nodes labeled 1-33 are staff and nodes labeled 34-107 are patients}.
\subfloat[RC measures without normalization]	
{\includegraphics[width = .5\textwidth]{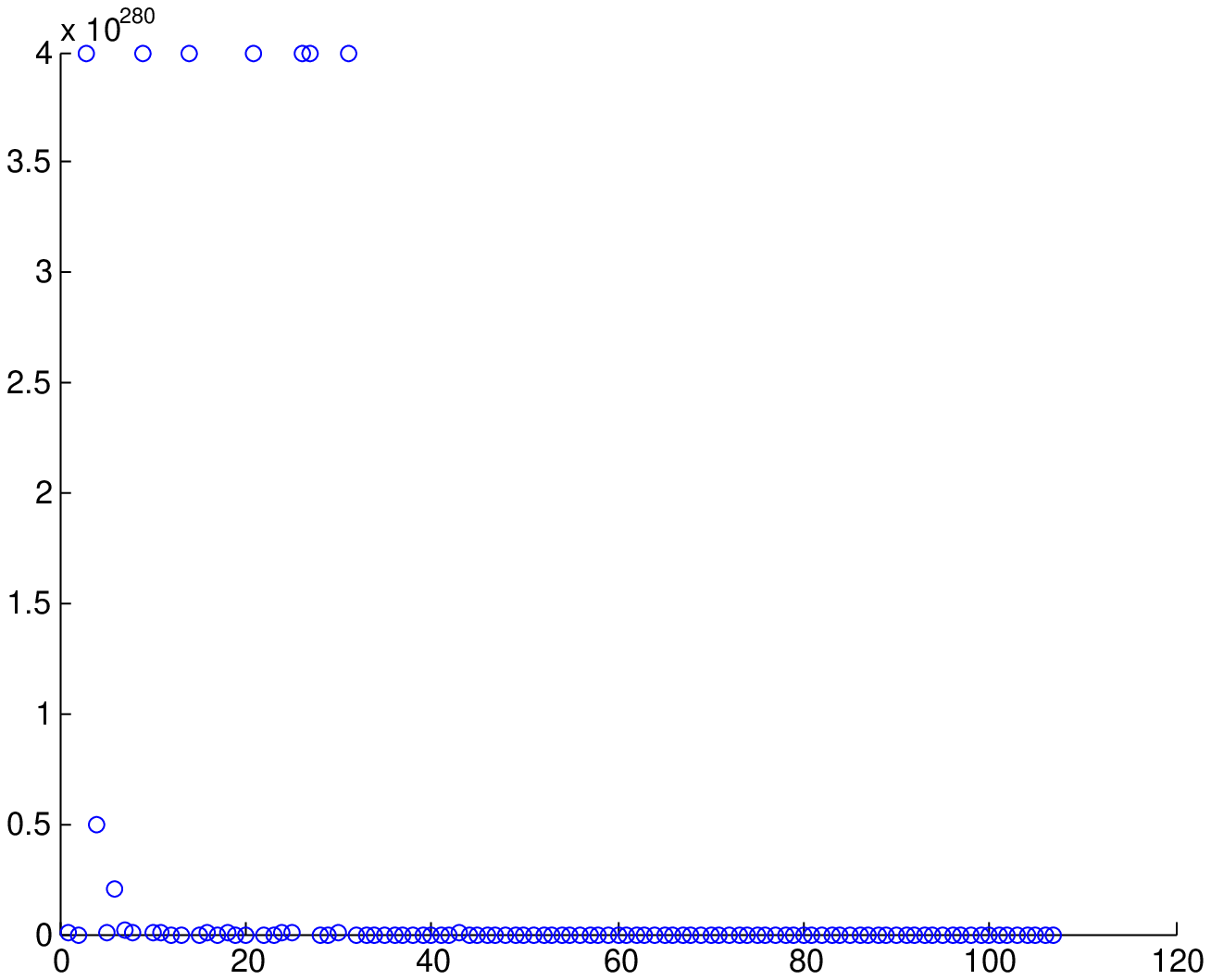}
\label{fig:RCnorm0}}
\subfloat[RC measures when adjacency matrices are aggregated into 20-min time frames]
{\includegraphics[width = .5\textwidth]{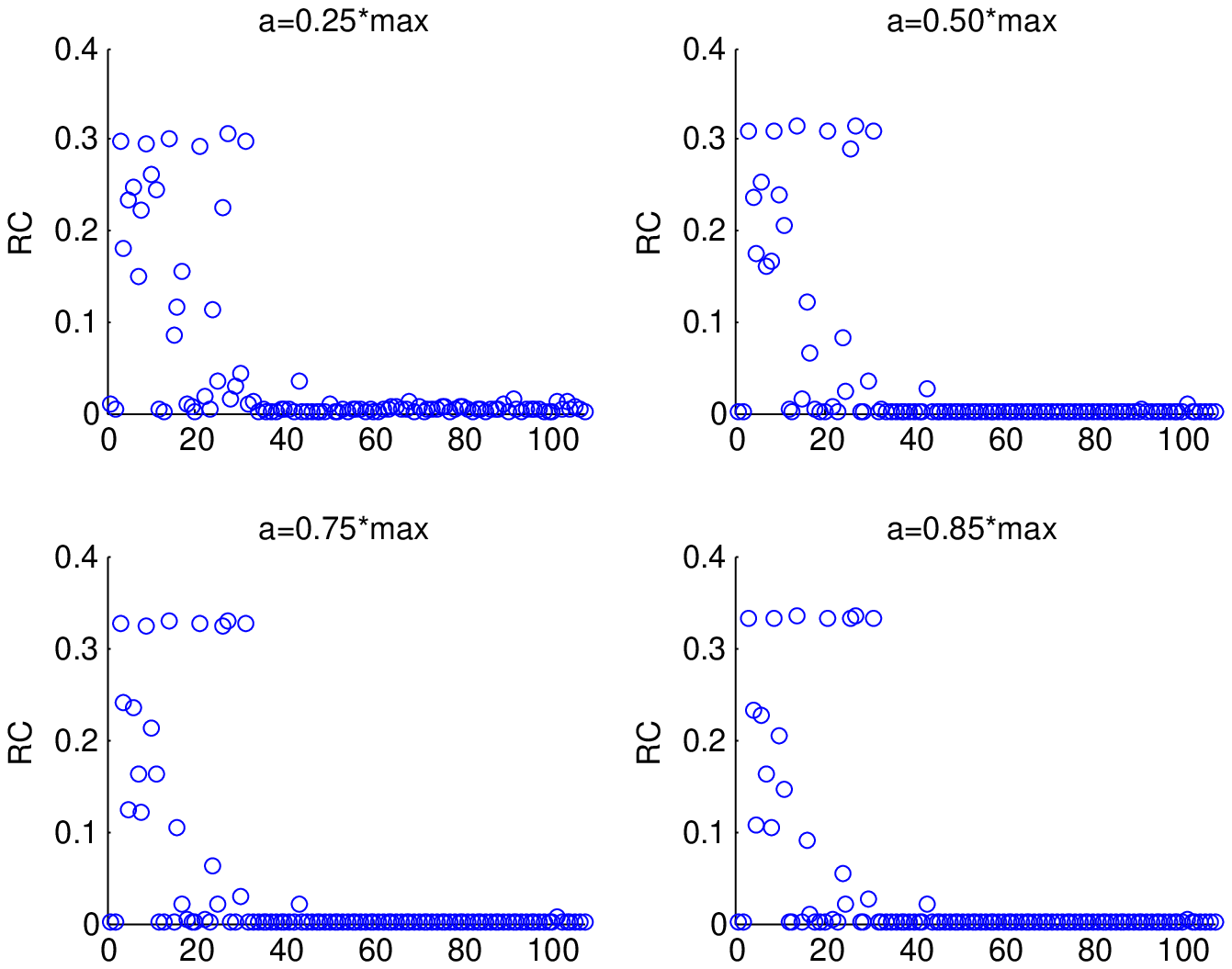}
\label{fig:RC20min}} 
\label{fig:RCconstraints}
\end{figure}

Figure \ref{fig:RCnorm0} plots the RC measures for $\alpha_1 = 0.25*\alpha_{\max}$ when $Q$ is not normalized. (There is overflow for $\alpha = \alpha_2, \alpha_3, \alpha_4$.) The same behavior is observed, ruling normalization out as an explanation for the highly skewed distribution of the RC measures. We conclude that from a `receiving' point of view, the same seven RN's are particularly distinct compared to the other nodes in the network, but are indistinguishable from each other. 

We point out that aggregating the base matrices $A^{[k]}$ into longer time frames (20 minutes instead of 10 seconds) resulted in better distinction among the top 20-30 RC nodes, as shown in Figure \ref{fig:RC20min}. Note, however, that the top seven nodes remain the same, and furthermore, high RC nodes are typically staff (nodes 1-33).  

\subsubsection{Comparison of node rankings}

\begin{figure}[h!]
\centering
\caption[Comparisons of node rankings for different values of $\alpha$]{Comparisons of node rankings for different values of $\alpha$. Rankings according to BC are shown in \ref{plot:ranksBC} and rankings according to RC are shown in \ref{plot:ranksRC}. The $(i,j)$-th position plots the rankings associated with $\alpha_i$ versus $\alpha_j$, for $i,j=1,\ldots,4$. Associated Pearson correlation coefficients are shown in \ref{ranks:pearson}.}
\subfloat[BC]		
{\label{plot:ranksBC}		\includegraphics[width = 3 in]{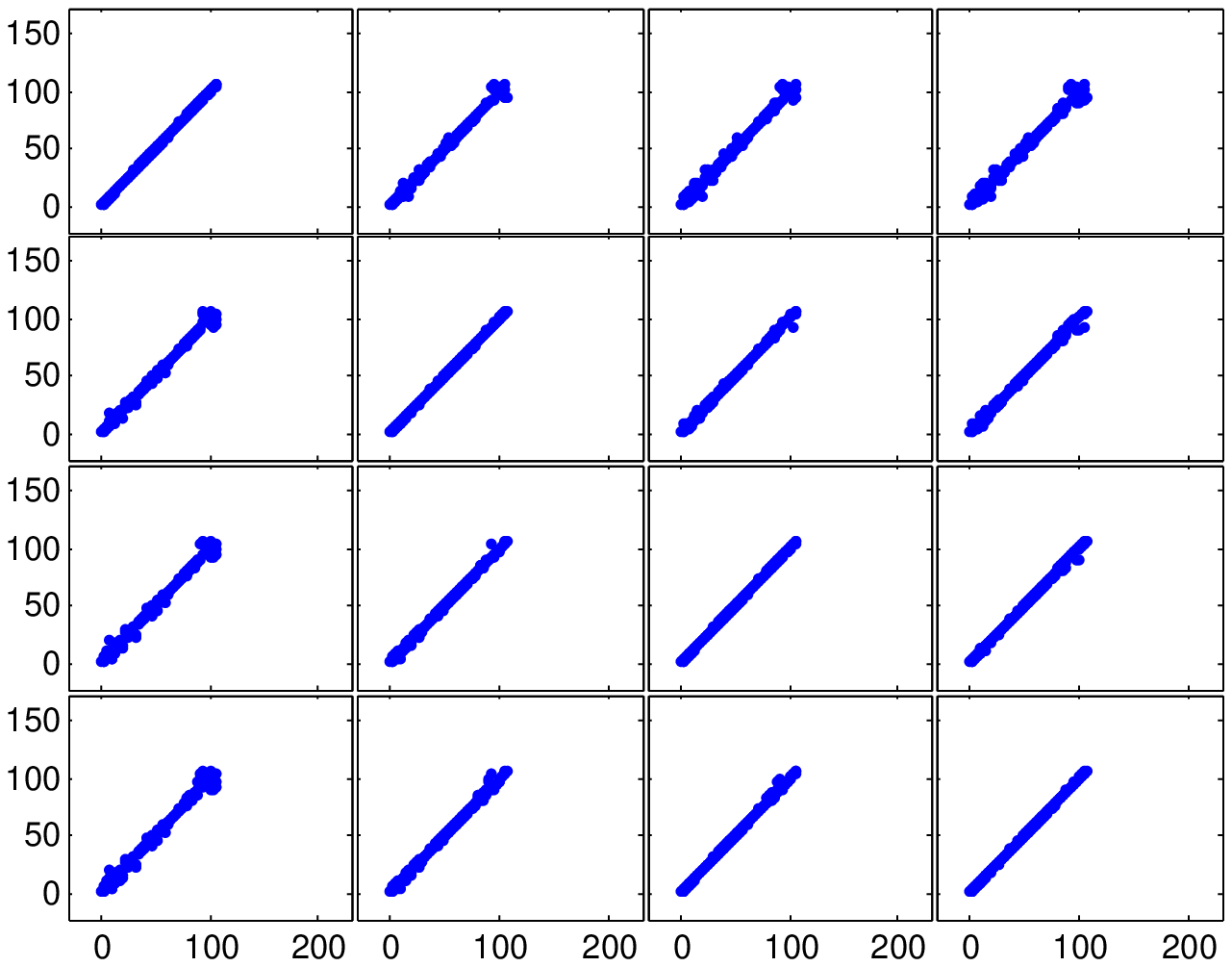}}
\subfloat[RC]
{\label{plot:ranksRC}
\includegraphics[width = 3 in]{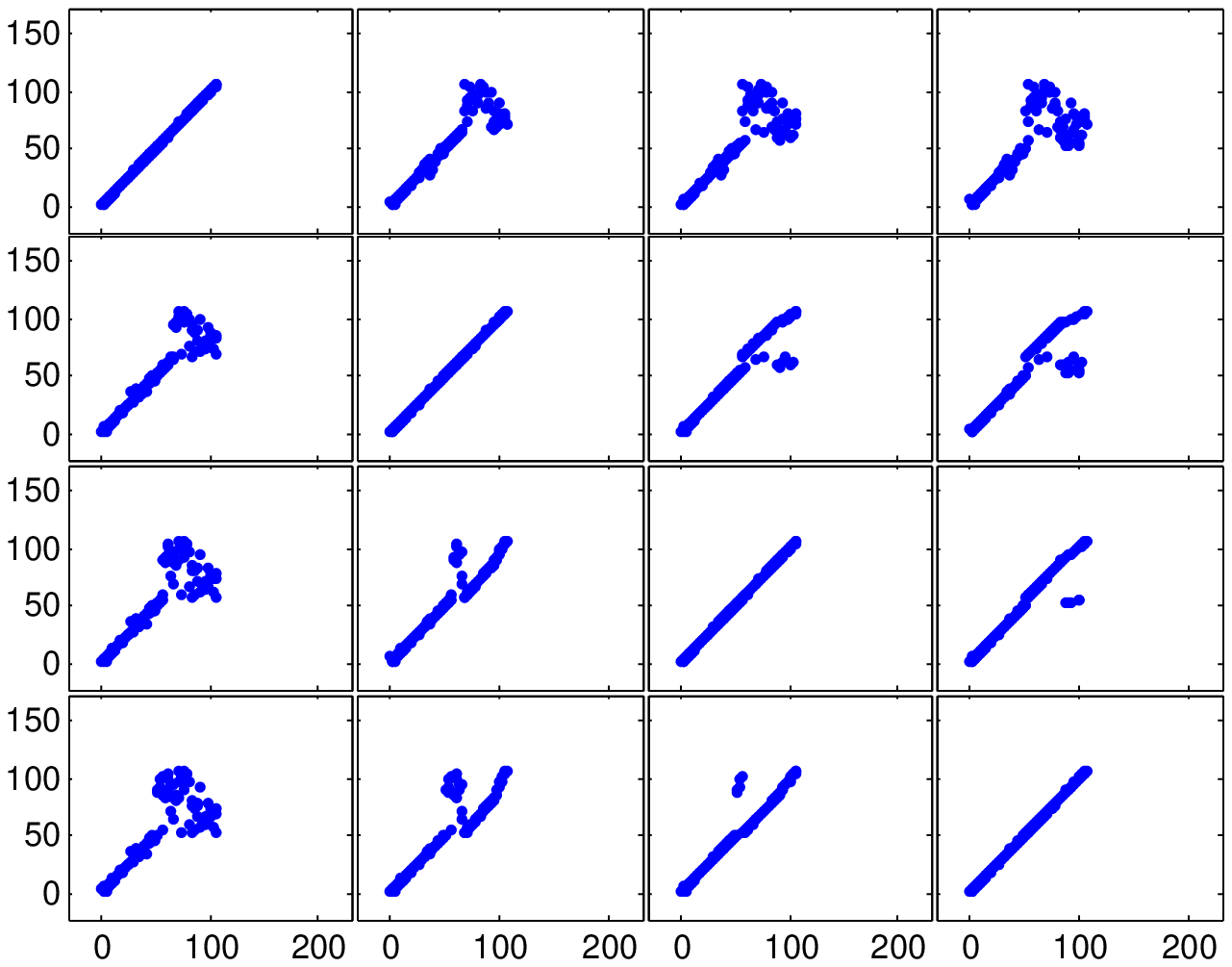}} \\
\subfloat[Associated Pearson correlation]
{\label{ranks:pearson}

{\arraycolsep=1.4pt\def\arraystretch{1.4}
\begin{tabular}{l|c|c|c|c|c|c}
\hline \hline 	&	$\alpha_1$ v $\alpha_2$	&	$\alpha_1$ v $\alpha_3$	&	$\alpha_1$ v $\alpha_4$	&	$\alpha_2$ v $\alpha_3$	&	$\alpha_2$ v $\alpha_4$	&	$\alpha_3$ v $\alpha_4$	\\	\hline
BC	&	0.9946	&	0.9913	&	0.9892	&	0.9982	&	0.9968	&	0.9986	\\	
RC	&	0.9164	&	0.8365	&	0.7765	&	0.9422	&	0.8919	&	0.9552	\\	\hline \hline
\end{tabular}}

}
\end{figure}

We consider the node rankings obtained based on BC and RC measures. Nodes are ranked from highest to lowest in descending order of the measures; the node with largest measure has rank 1. For each value of $\alpha_i$ for $i=1,\ldots,4$, we have a corresponding list of rankings $l_{\alpha_i}$ where
\[l_{\alpha_i}(k):= \text{ranking of node }k \text{ when } \alpha = \alpha_i. \]
The $(i,j)$-th position in Figure \ref{plot:ranksBC} plots $l_{\alpha_i}$ versus $l_{\alpha_j}$ where the rankings are based on BC measures. In Figure \ref{plot:ranksRC} we plot lists of rankings based on RC measures. The Pearson correlations corresponding to these plots are shown in Table \ref{ranks:pearson}. 

\begin{figure}[h!]
\caption[Spaghetti plots comparing node rankings for different values of $\alpha$]{Spaghetti plots display a line for each node connecting the rankings obtained for the different values of $\alpha$. The closer the line is to horizontal, the more similar the rankings are to each other. Rankings based on aggregate degree (AD) are labeled $\alpha=0$. }
\subfloat[BC]
{\includegraphics[width = 3 in]{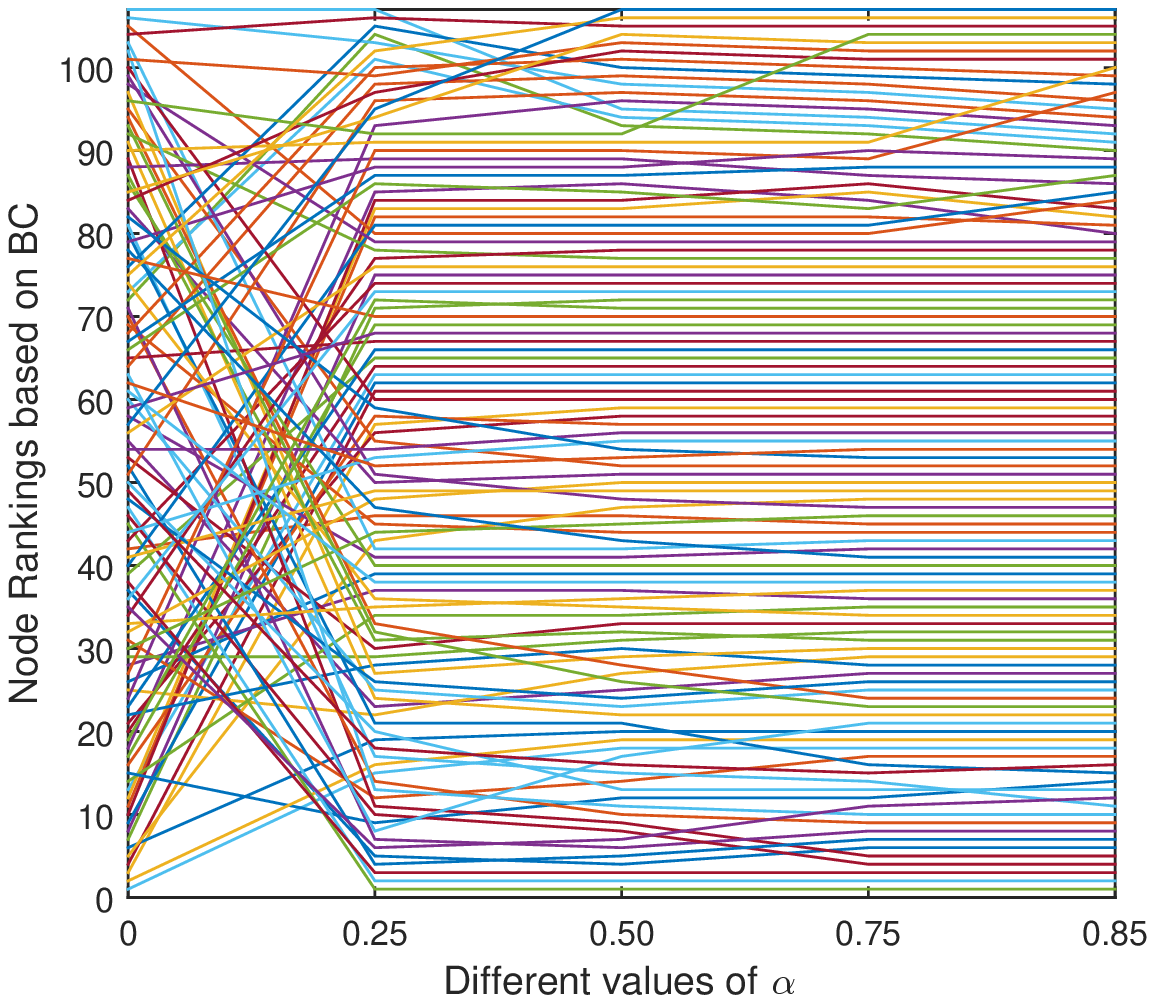}}
\subfloat[RC]
{\includegraphics[width = 3 in]{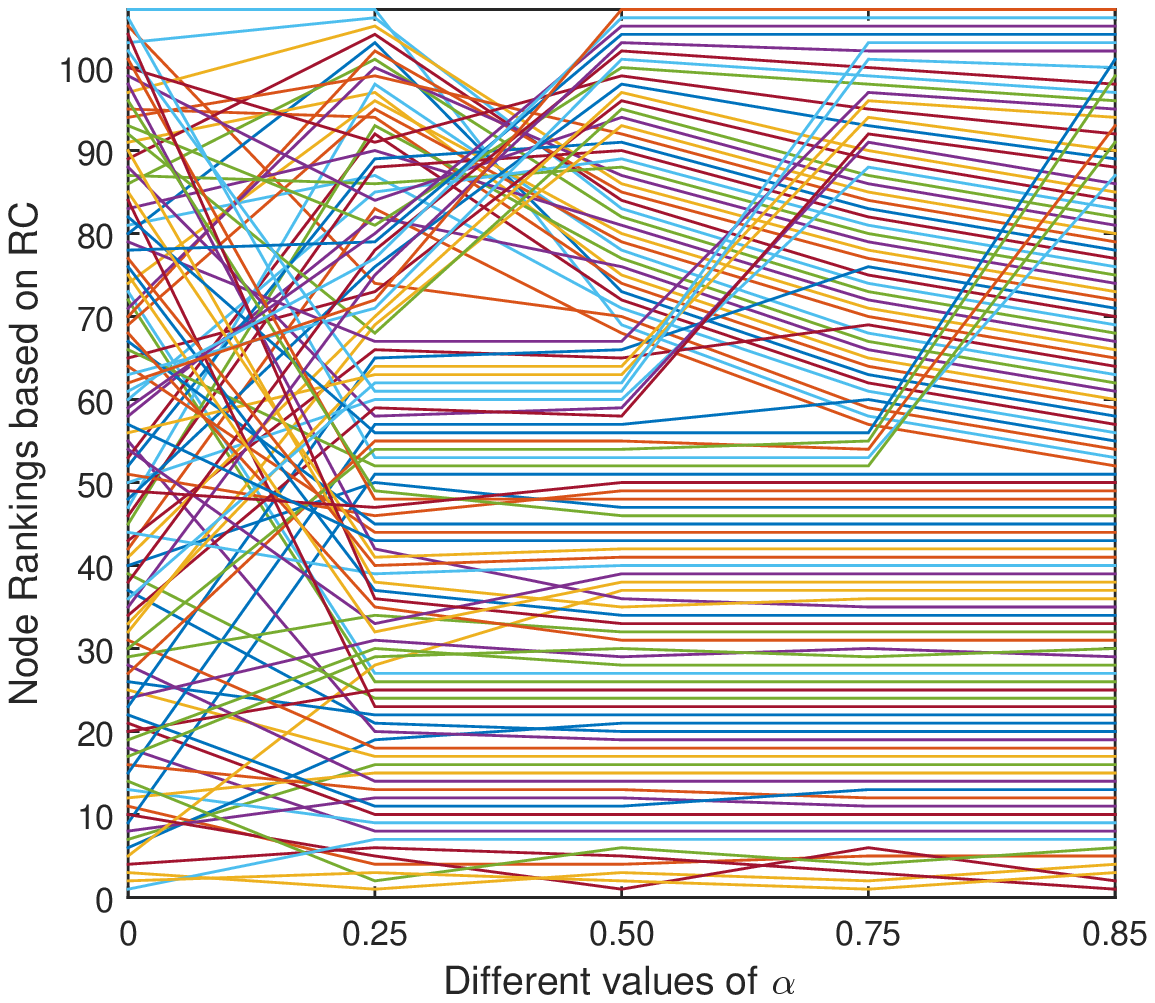}}
\label{spaghettiPlots}
\end{figure}

In Figure \ref{spaghettiPlots}, we see that the rankings obtained for $\alpha = \alpha_1,\alpha_2,\alpha_3,\alpha_4$, are relatively robust and furthermore, the chosen values for $\alpha$ are far enough from the limit $\alpha \rightarrow 0$, so that the rankings are significantly different from those based on AD. Noisy behavior for low-ranked RC nodes is probably due to the small values of the RC measures: small changes can lead to drastic changes in rankings among low-ranked nodes. 

\begin{figure}[h!]
\centering
\caption[Comparison of rankings between staff and patients]{Comparison of rankings between staff and patients. We report the average ranks over $\alpha_1,\ldots,\alpha_4$. On each box, the horizontal line is the median, the edges of the box are the 25th and 75th percentiles, the whiskers extend to the most extreme data points not considered outliers, and outliers are plotted individually.}
\subfloat[Mean BC ranks]{\includegraphics[width =0.5\textwidth]{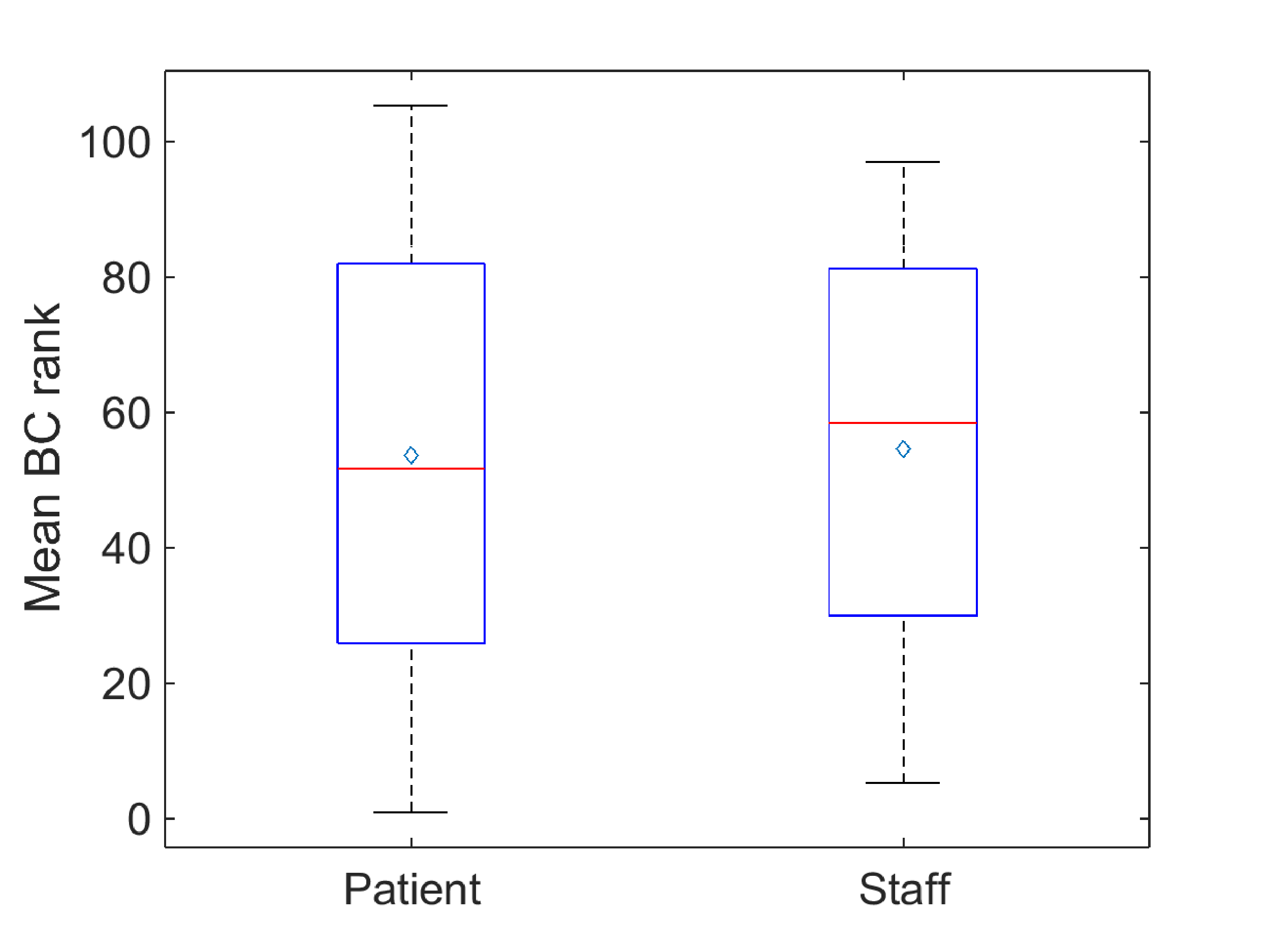}}
\subfloat[Mean RC ranks]{\includegraphics[width =0.5\textwidth]{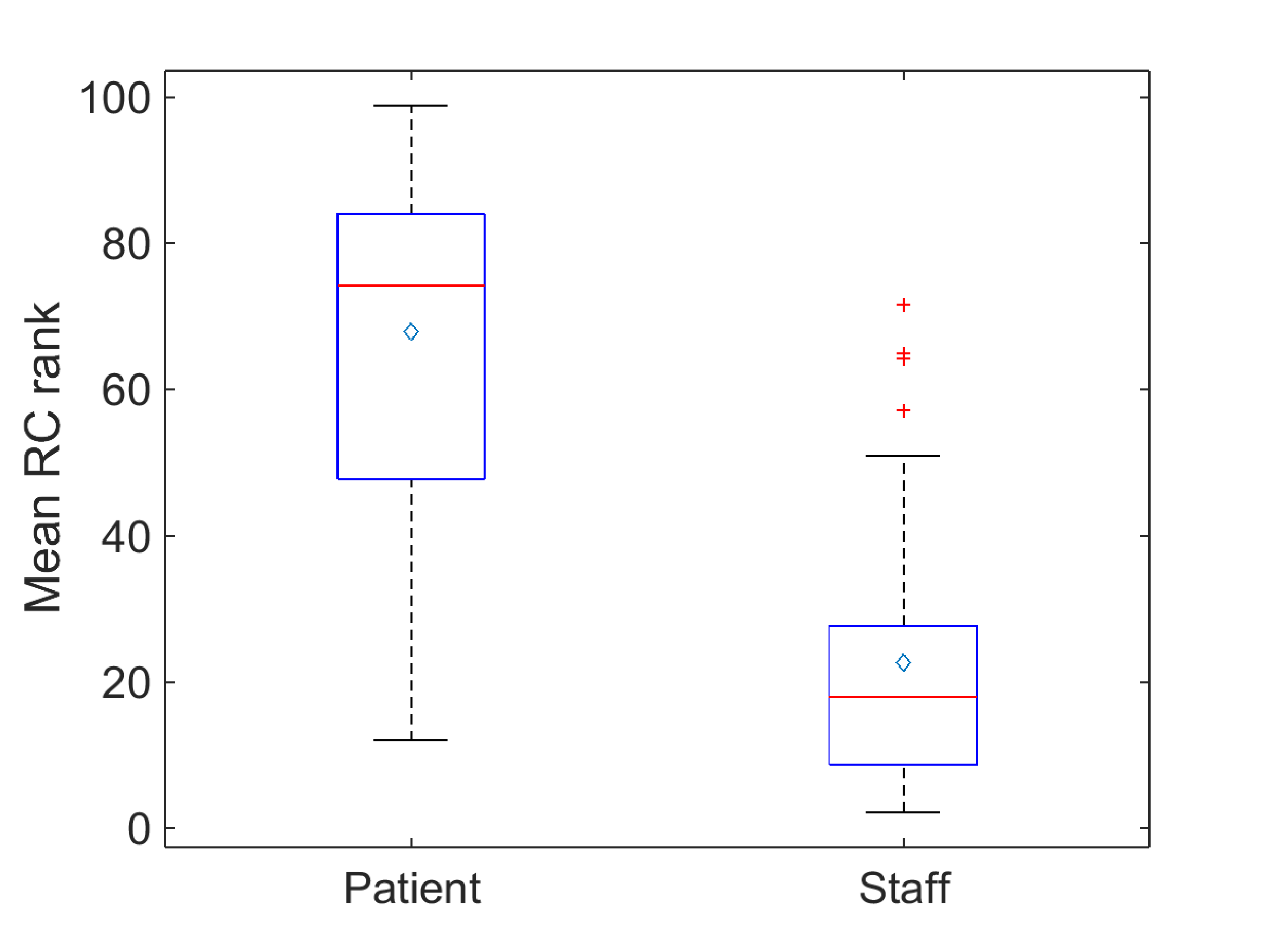}}
\label{boxplots: staff v pat}
\end{figure}

From Figure \ref{boxplots: staff v pat} we also see that patients tend to be, on average, slightly better broadcasters than staff. As mentioned, the fact that high receivers are predominantly staff is not surprising, given their roles in the ED. Staff are well-placed to be at the receiving ends of dynamic walks.   

\subsubsection{Comparison of nodes in ranked order} 

Consider lists of nodes in ranked order. Explicitly, for each value of $\alpha_i$ for $i = 1,\ldots,4$, we have a corresponding list of nodes $t_{\alpha_i}$ where
\[ t_{\alpha_i}(k) := \text{node that has rank }k \text{ when } \alpha = \alpha_i. \]
We compute Kendall correlation and intersection distance $isim$ \cite{isim} as quantitative ways to assess similarity between the lists. Small values of $isim \in [0,1]$ are indicative of strong similarity between lists. These are shown in Table \ref{table: kcorr} and Table \ref{table: isim}. 


\begin{table}[h!]
\centering
\footnotesize
\arraycolsep=1.4pt\def\arraystretch{1.4}
\caption[Kendall correlation between lists of nodes in ranked order]{Kendall correlation between lists of nodes in ranked order.}
\label{table: kcorr}

\begin{tabular}{l|c|c|c|c|c|c}
\hline \hline 	&	$a_1$ v $a_2$	&	$a_1$ v $a_3$	&	$a_1$ v $a_4$	&	$a_2$ v $a_3$	&	$a_2$ v $a_4$	&	$a_3$ v $a_4$	\\	\hline
BC	&	0.2449	&	0.1963	&	0.1825	&	0.5204	&	0.4579	&	0.7366	\\	
RC	&	0.3871	&	0.2918	&	0.2721	&	0.7411	&	0.6805	&	0.8173	\\	\hline \hline
\end{tabular}
 \\~\\
\caption[Intersection distance ($isim$) between lists of nodes in ranked order]{Intersection distance ($isim$) between lists of nodes in ranked order. Values of $isim$ close to $0$ are indicative of strong similarity between lists.}

\begin{tabular}{l|c|c|c|c|c|c}
\hline \hline 	&	$\alpha_1$ v $\alpha_2$	&	$\alpha_1$ v $\alpha_3$	&	$\alpha_1$ v $\alpha_4$	&	$\alpha_2$ v $\alpha_3$	&	$\alpha_2$ v $\alpha_4$	&	$\alpha_3$ v $\alpha_4$	\\	\hline
BC	&	0.0330	&	0.0518 & 0.0561 &	0.0260 &	0.0305 &	0.0060 \\	
RC	&	0.0749 &	0.0999 &	0.1286 &	0.0610 & 0.0784 &	0.0525 \\	\hline \hline
\end{tabular}

\label{table: isim}
\end{table}


\subsection{Convergence to AD}
Recall that as $\alpha \rightarrow 0$, both BC and RC rankings should converge to AD rankings (see Section \ref{section:Limit}). We present the results for small values of $\alpha$ approaching zero, and in Table \ref{table:convergenceAD} we see that according to various measures, BC and RC rankings do indeed approach AD rankings\footnote{In Table 
\ref{table:convergenceAD} we observe an initial drop in Kendall correlation between nodes ranked according to AD versus RC (see \% $\alpha_{\max} = 0.005$). Since the  correlation values are at the low end (0.25, 0.20), this anomaly is probably due to a small change in the number of discordant versus concordant pairs, and is in itself not a significant departure from the overall trend of convergence.}. This provides added assurance that in spite of the fact that the computed BC and RC measures are numerically tiny, the rankings obtained are nonetheless correct. 

\begin{table}[h!]
\centering
\footnotesize
\arraycolsep=1.4pt\def\arraystretch{1.4}
\caption[Comparisons of AD, BC, RC node rankings for $\alpha$ approaching zero]{Convergence to AD: Comparisons of AD, BC, RC node rankings for $\alpha$ approaching zero. Pearson correlation $(corr)$ compares lists of node rankings. Intersection distance $(isim)$ and Kendall correlation $(kcorr)$ compare lists of nodes in ranked order. We also report the number of nodes in common among the top 10 and top 5.}
\begin{tabular}{l|c|c|c|c|c|c|c}
\hline\hline $\%$ of $ \alpha_{\max}$	&	0.01	&	0.005	&	$10^{-4}$	&	$10^{-5}$	&	$10^{-6}$	&	$10^{-7}$	&	$10^{-8}$	\\\hline
corr(AD,RC)	&	0.74735	&	0.932258	&	0.998119	&	0.999951	&	0.999990	&	0.999990	&	0.999990	\\
corr(AD,BC)	&	0.636092	&	0.80269	&	0.997962	&	0.999951	&	0.999990	&	0.999990	&	0.999990	\\
corr(RC,BC)	&	0.270381	&	0.675454	&	0.99615	&	0.999892	&	0.999980	&	0.999980	&	0.999980	\\
Top 10 intersection	&	4	&	4	&	7	&	10	&	10	&	10	&	10	\\
Top 5 intersection	&	2	&	3	&	3	&	5	&	5	&	5	&	5	\\
isim(AD,RC)	&	0.219576	&	0.135918	&	0.034806	&	0.002072	&	0.000091	&	0.000091	&	0.000091	\\
isim(AD,BC)	&	0.250046	&	0.178938	&	0.028027	&	0.001529	&	0.000123	&	0.000123	&	0.000123	\\
isim(RC,BC)	&	0.364357	&	0.250965	&	0.051588	&	0.003602	&	0.000214	&	0.000214	&	0.000214	\\
kcorr(AD,RC)	&	0.251984	&	0.19873	&	0.435373	&	0.877623	&	0.965086	&	0.965086	&	0.965086	\\
kcorr(AD,BC)	&	0.115147	&	0.343326	&	0.348968	&	0.986598	&	0.989773	&	0.989773	&	0.989773	\\
kcorr(RC,BC)	&	0.119732	&	0.253747	&	0.299242	&	0.864927	&	0.955563	&	0.955563	&	0.955563	\\\hline\hline

\end{tabular}

\label{table:convergenceAD}
\end{table}

\begin{figure}[h!]
\centering
\caption[Comparisons of AD, BC, RC node rankings for $\alpha$ approaching zero]{Convergence to AD: Comparisons of AD, BC and RC node rankings for $\alpha$ approaching zero.}
\subfloat[$0.01*\alpha_{\max}$]{\includegraphics[width = 3 in]{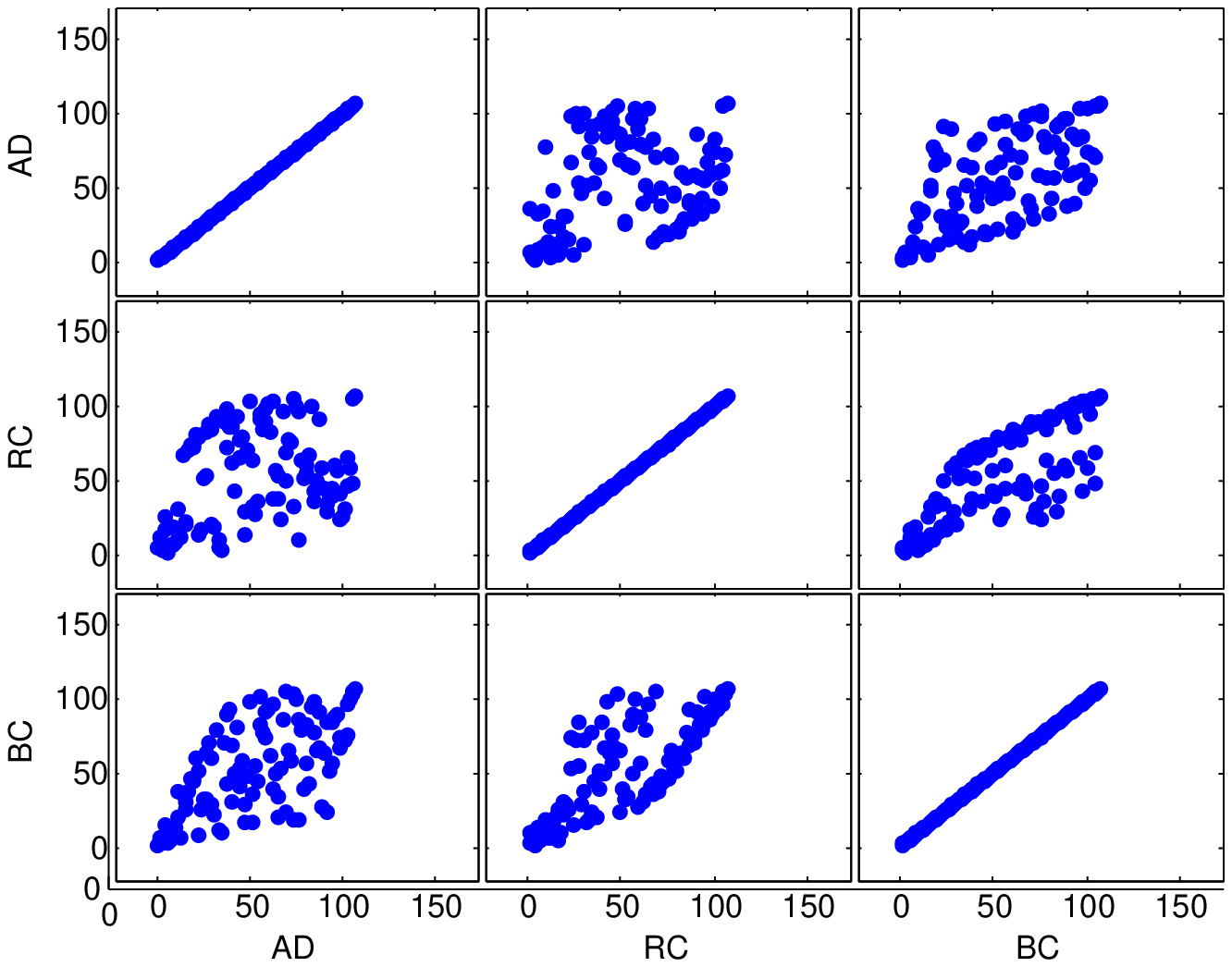}}
\subfloat[$0.005*\alpha_{\max}$]{\includegraphics[width = 3 in]{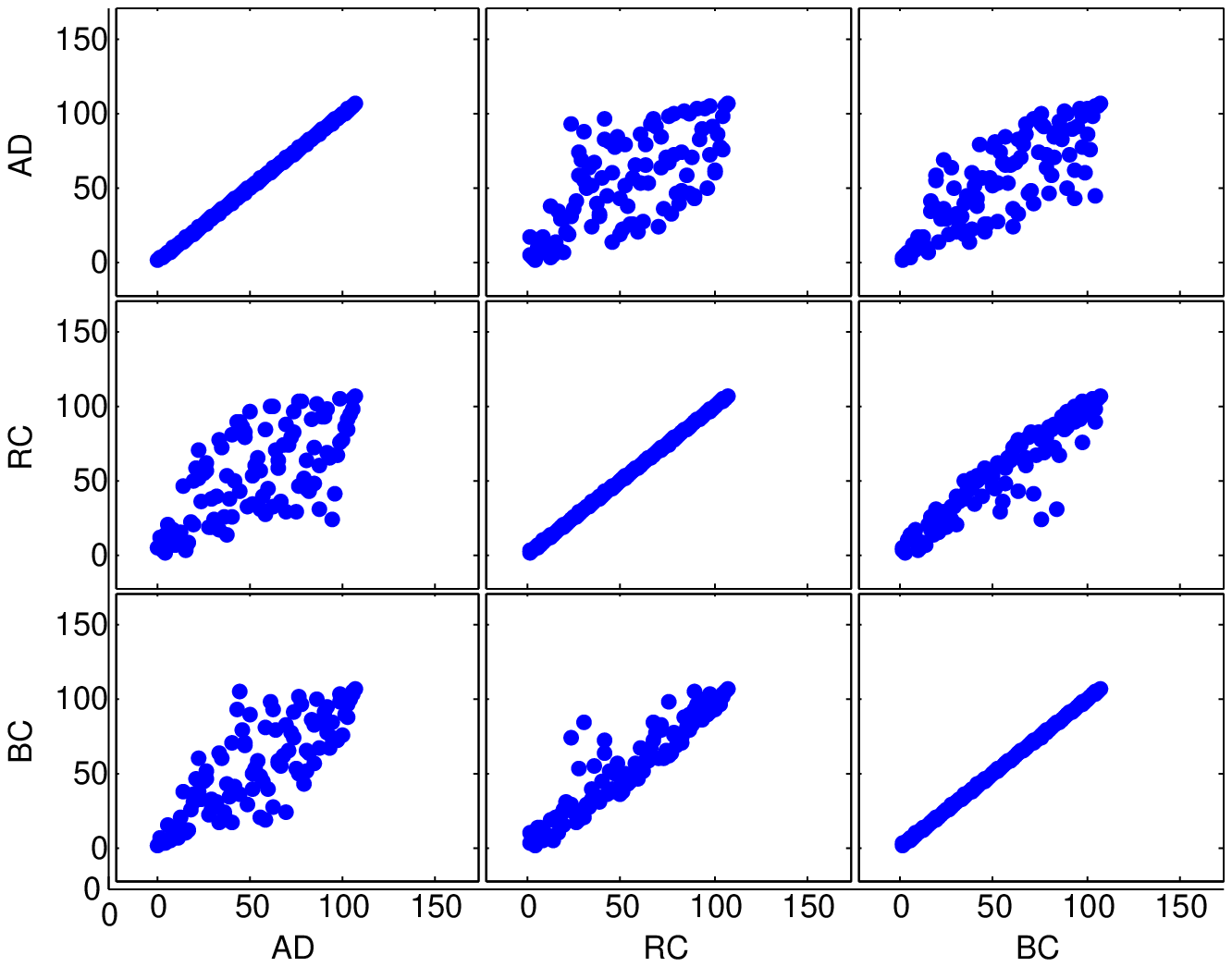}} \\
\subfloat[$0.001*\alpha_{\max}$]{\includegraphics[width = 3 in]{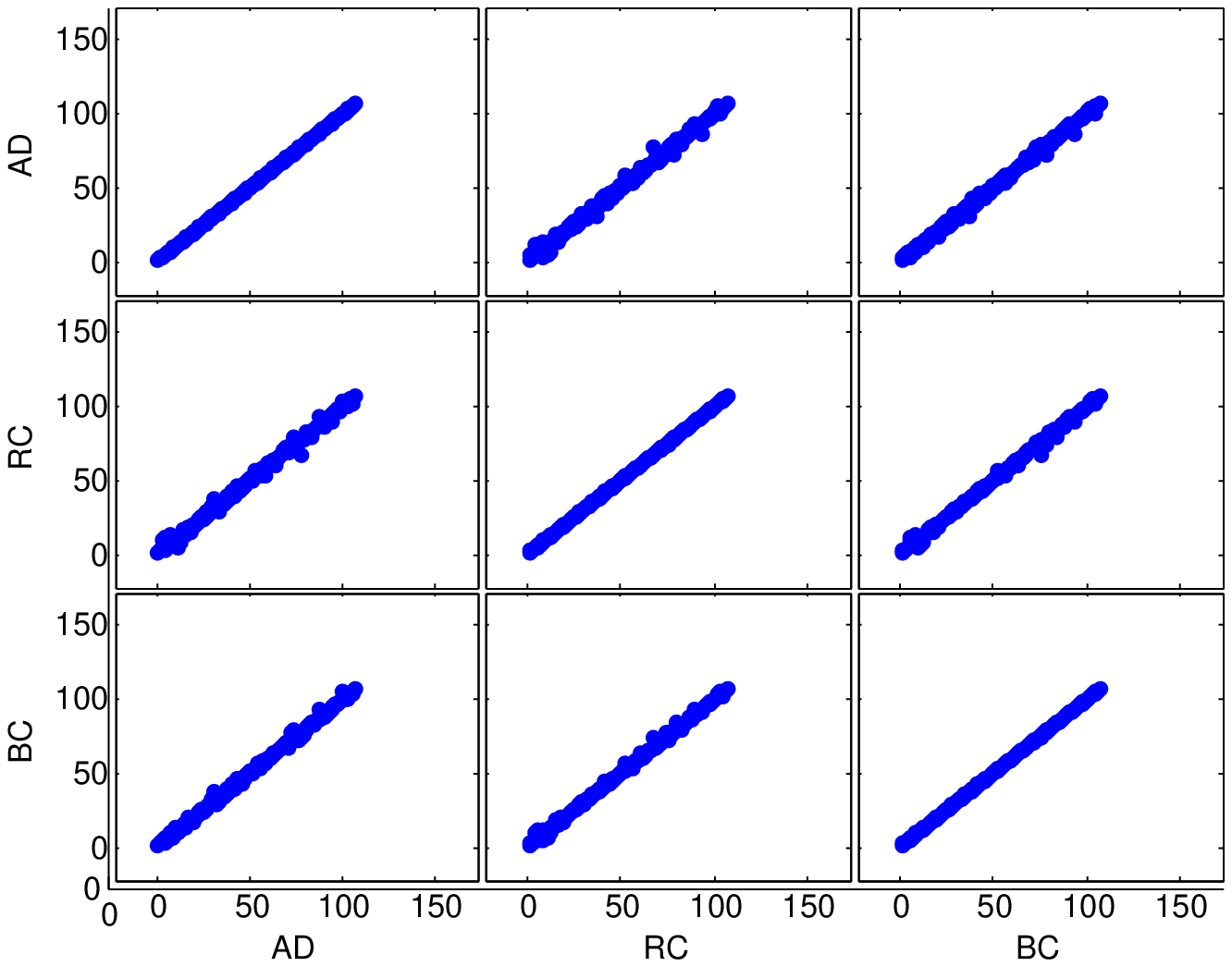}}
\subfloat[$0.0001*\alpha_{\max}$]{\includegraphics[width = 3 in]{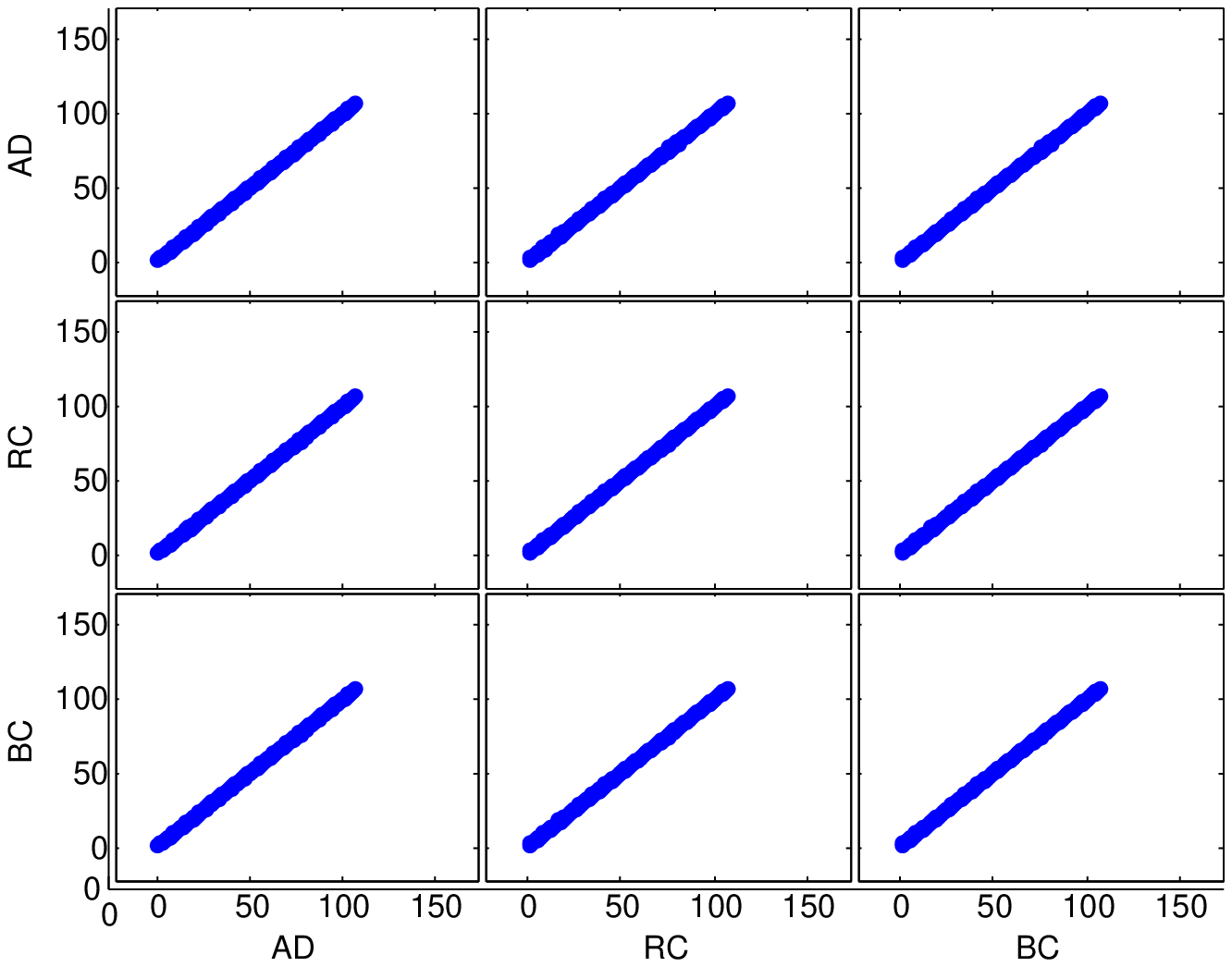}} 
\end{figure}

\begin{figure}[h!]
\centering
\caption[Spaghetti plots of node rankings for $\alpha$ approaching zero]{Convergence to AD: Spaghetti plots of node rankings based on AD, BC and RC for different values of $\alpha$ approaching zero. The closer the lines are to horizontal, the more similar the AD, BC and RC-based rankings are to each other. }
\subfloat[$0.01*\alpha_{\max}$]{\includegraphics[width = 3 in]{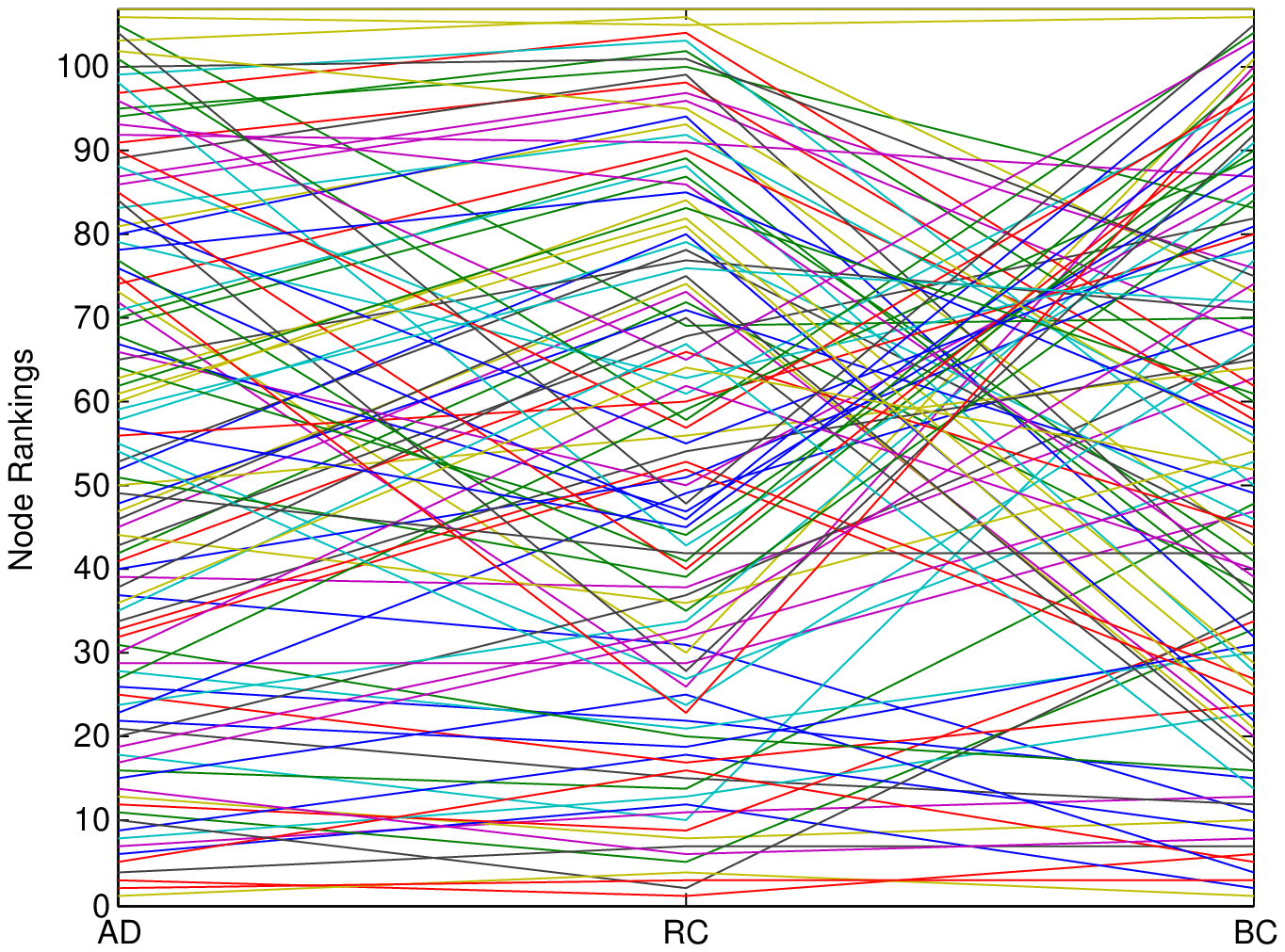}}
\subfloat[$0.005*\alpha_{\max}$]{\includegraphics[width = 3 in]{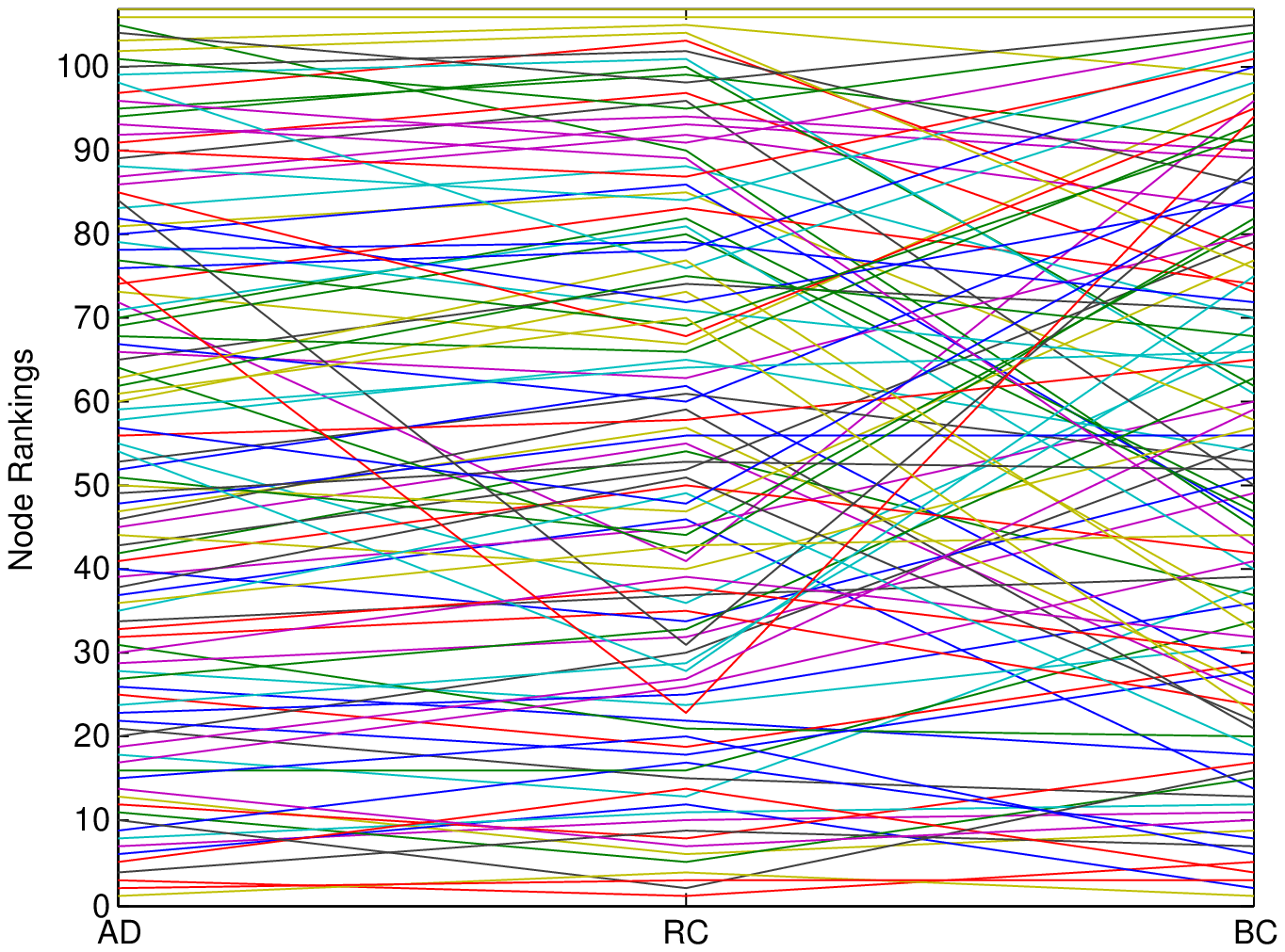}} \\
\subfloat[$0.001*\alpha_{\max}$]{\includegraphics[width = 3 in]{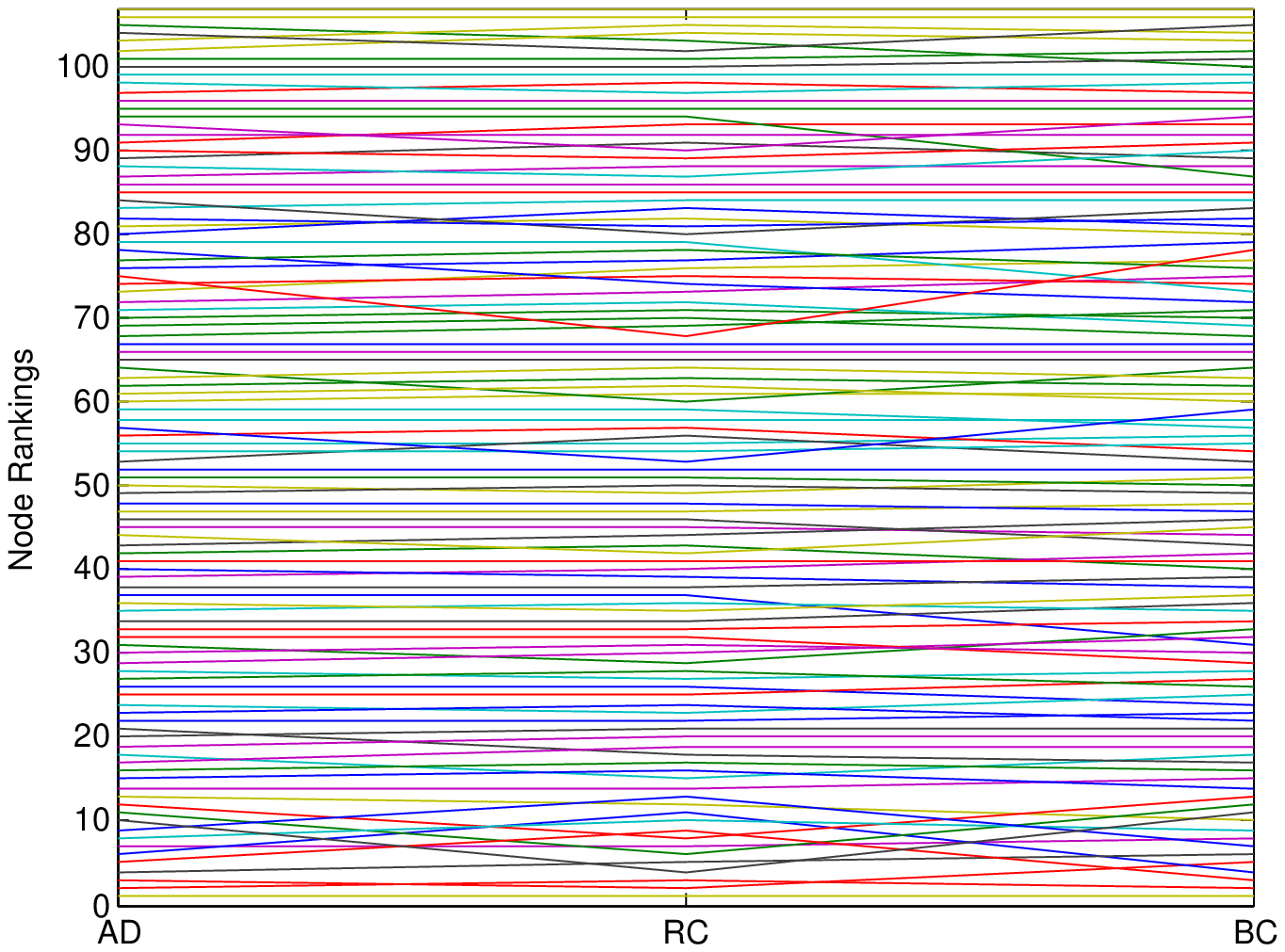}}
\subfloat[$0.0001*\alpha_{\max}$]{\includegraphics[width = 3 in]{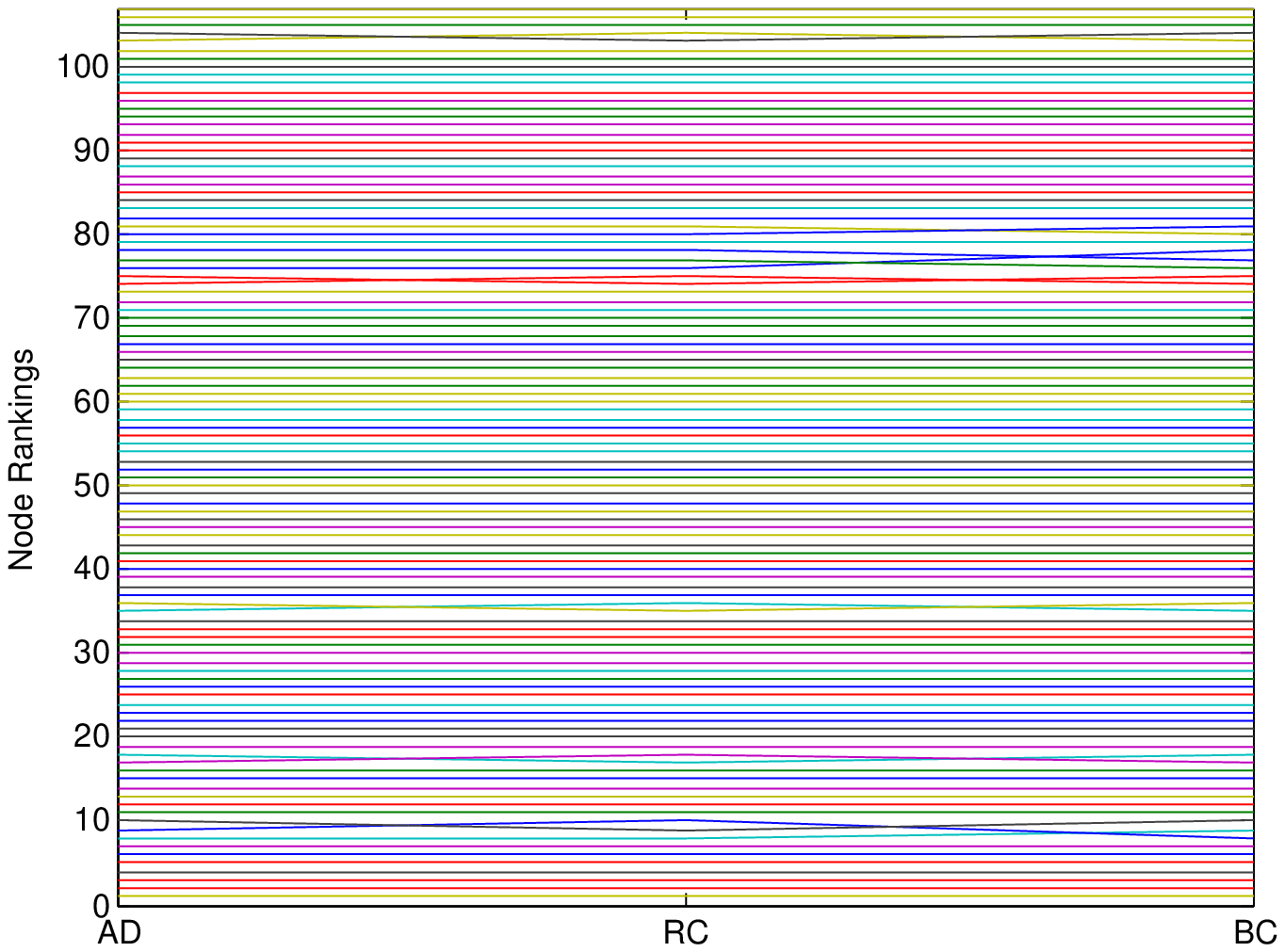}} 
\end{figure}

\subsection{Dynamic communicability based on the matrix exponential}
\label{Section: Exp}

\begin{figure}[h!]
\centering
\caption[Dynamic communicability based on the matrix exponential (shift 1)]{Dynamic communicability based on the matrix exponential (shift 1). There is one data point per node; the horizontal axis is the node ID label. Nodes labeled 1-33 are staff and nodes labeled 34-107 are staff.}
\subfloat[$BC$]
{\includegraphics[width = 3 in]{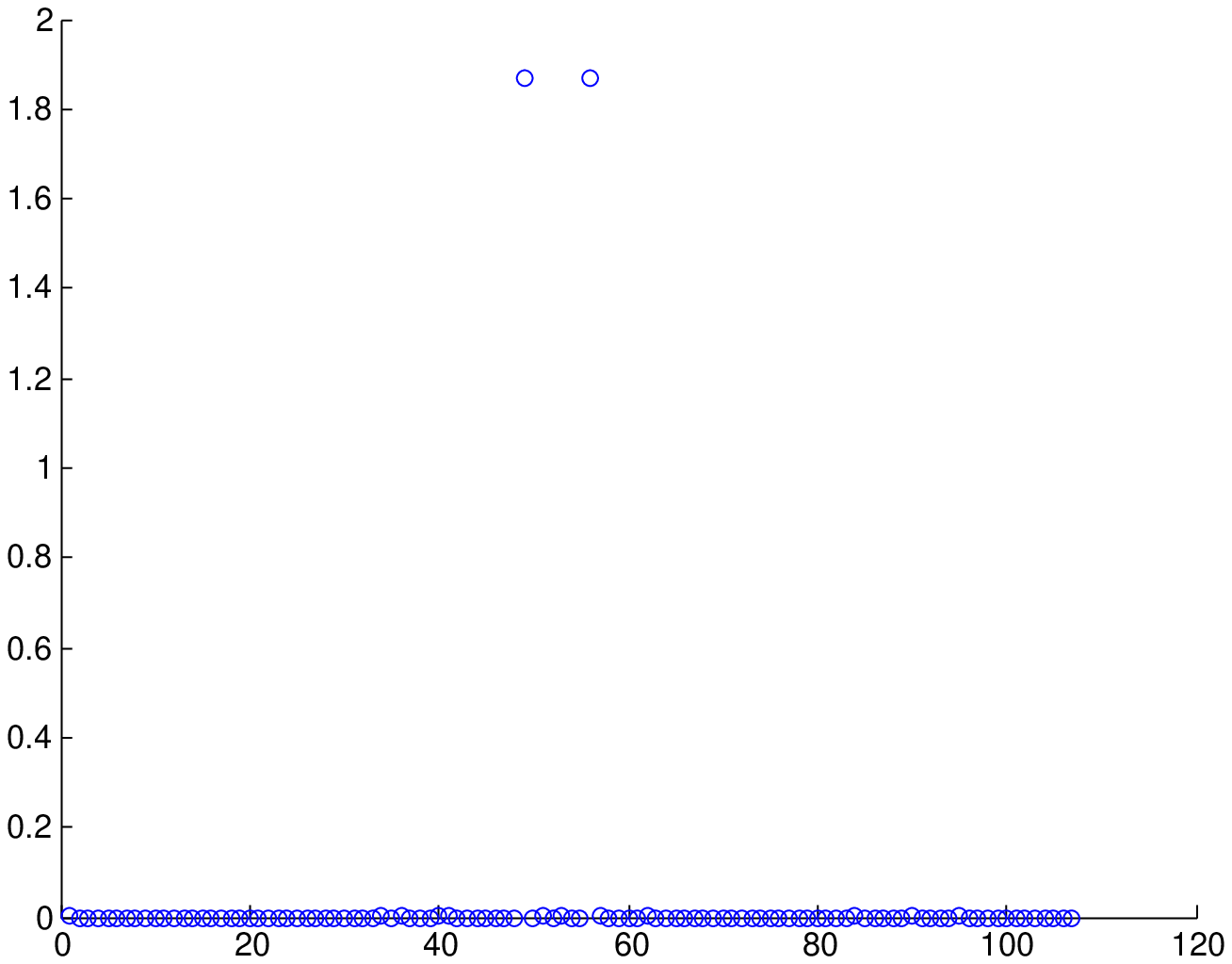}}
\subfloat[$RC$]
{\includegraphics[width = 3 in]{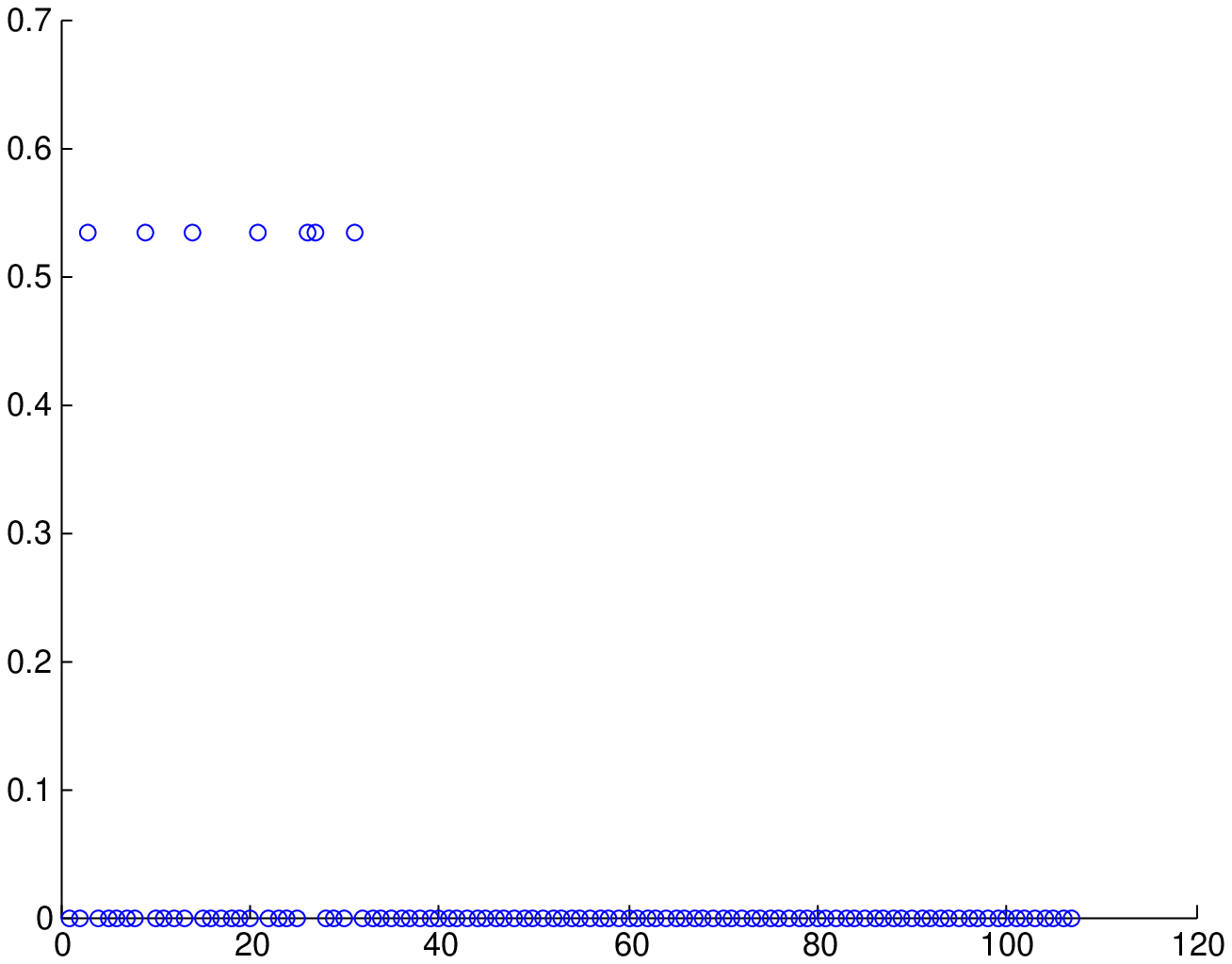}}
\label{Fig: Qexp}
\end{figure}

\begin{figure}[h!]
\centering
\caption{Comparison of node rankings between dynamic communicability based on the resolvent versus the matrix exponential.}
\subfloat[$BC$]
{\includegraphics[width = 3 in]{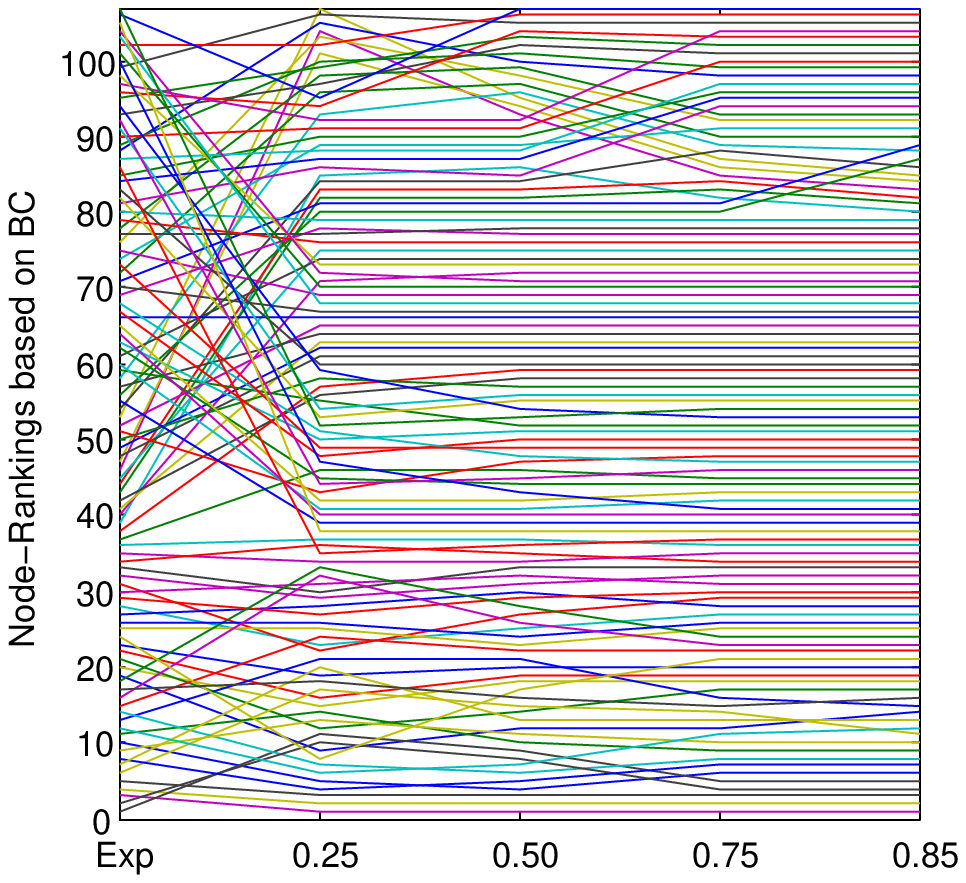}}
\subfloat[$RC$]
{\includegraphics[width = 3 in]{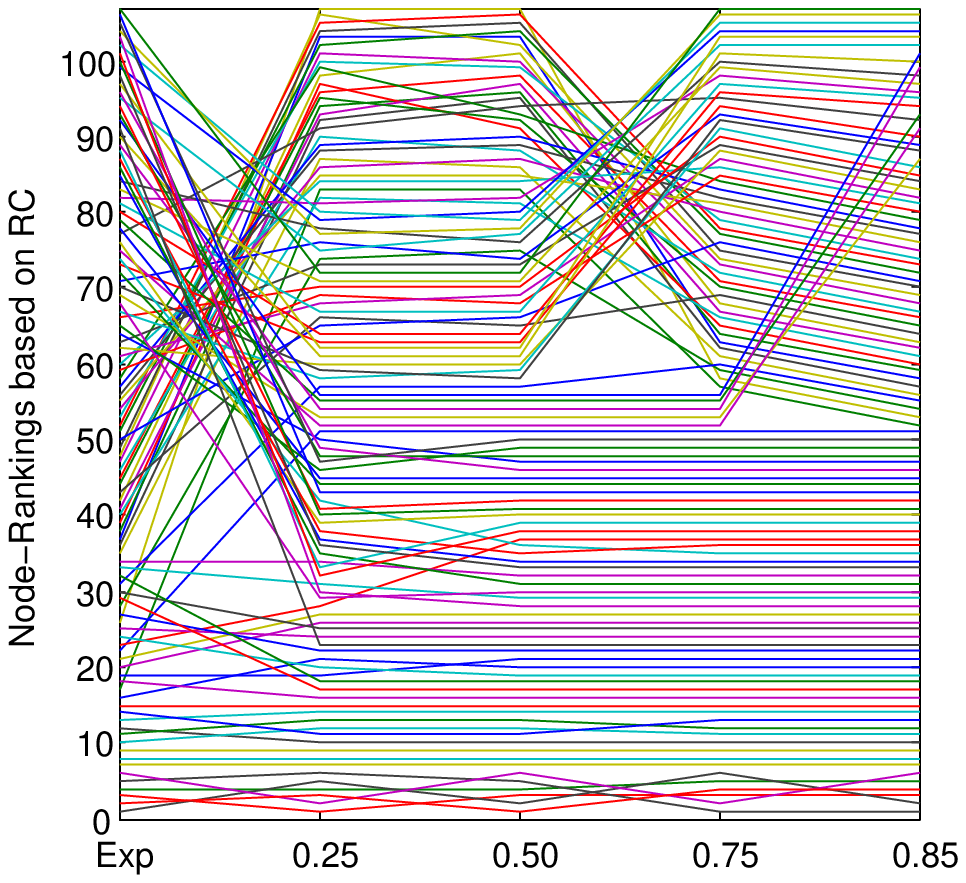}}
\label{Fig: spaghetti-Exp}
\end{figure}

We compute a version of dynamic communicability based on the matrix exponential as defined in Eq. (\ref{Eq: Qexp}). BC and RC measures obtained are shown in Figure \ref{Fig: Qexp}. We see that $BC$ measures based on the matrix exponential are less able to distinguish between the top $20$ nodes. In Figure \ref{Fig: spaghetti-Exp} the resulting rankings are shown in comparison to the resolvent-based formulation of $Q$. The rankings obtained appear to be fairly similar, differing mostly in the low-ranking nodes. 

\section{Infection model}\label{InfModel}
We use a stochastic approach to model the spread of disease in a dynamic network, where the length of contact time between nodes is explicitly used to compute the probability of infection. The dynamic nature of the contacts is exploited thereby distinguishing our model from traditional epidemic processes on static networks \cite{goltsev,newman-epi, vespignani}. Without the temporal dimension, edges representing potential disease-spreading contacts are considered present and unchanging over time. Consequently, assuming there is no recovery, it is only a matter of time before everyone in (connected components of) the network is infected. However, in reality, contacts themselves form and dissolve over time. By incorporating the dynamic nature of contacts into the infection model, we hope to paint a more realistic picture of how contagion spreads on a network \cite{bansal,Volz+Meyers}.  

We emphasize that the contagion process on the network is completely independent of the notion of dynamic communicability as discussed in Section \ref{DC}. 

Parameters for infection are chosen based on rates observed for influenza \cite{airline_flu, flu,Salathe}. Within the framework of a limited observation period, we assume that infected nodes are immediately infectious. In addition, since recovery is not physically feasible within this time frame, the Susceptible-Infected (SI) model is adequate, and models the early phase of an outbreak. 

For each simulation, there is only one initial source of infection, which is infectious upon its arrival in the ED. Our aim is to associate a measure of virulence with each node. This is in contrast to the work done in \cite{Rocha-SexNetwork} where the source of infection is chosen randomly and proportional to the number of contacts. 

We assume that infection between Susceptible-Infected pairs is a Poisson process, which is, in particular, independent and memoryless. Consequently, the time to infection, $X$, follows an exponential distribution. We write
$X \sim \text{Exp}(\lambda^*)$. The parameter $\lambda^*$ is chosen to satisfy
\[\text{Pr}(X\leq 60 \text{ sec} ~|~\lambda^*)= \int_0^{60} \lambda^* e^{-\lambda^* t}\,dt = 1 - e^{-60\lambda^*} = 0.009,\]
where the value of $0.009$ is chosen based on an approximated attack rate observed in an outbreak of influenza aboard a commercial airliner \cite{airline_flu, flu,Salathe}. Solving for $\lambda^*$, we obtain $\lambda^* \approx 1.5\times 10^{-4}$. Note that $1/\lambda^* \approx 664$ seconds is interpreted as the average time to infection. Suppose nodes $i$ and $j$ are in contact for $t_{ij}$ seconds, then the probability of disease transmission is given by 
\[p_{ij} = \int_0^{t_{ij}} \lambda^* e^{-\lambda^* t}\,dt.\]
In order to determine if infection takes place, a random number $u\in \text{Unif}(0,1)$ is generated. If $u<p_{ij}$, infection occurs, otherwise, infection does not occur. The rationale here is that for `large' $p_{ij}$, we would like infection to occur as much as possible. Note that $p_{ij}$ is computed based on the entire length of uninterrupted contact time between nodes $i$ and $j$. Infection, and therefore further spreading potential, occurs not at the end of the contact period $t_{ij}$, but at the time-step when $p_{ij}$ exceeds $u$. Note that if the pair $(i,j)$ comes into contact multiple times over the course of study, each contiguous contact period is treated independently.

\section{Simulation results}\label{SimResults}
We present the results of the infection simulations performed on the temporal network of Shift 1. We consider each node in turn as the initial source of infection. For each initial source, we repeat the simulations $N=1000$ times. We consider the total number of infections, or final epidemic size (EPI), that occur during the entire shift. Observed distributions of EPI per seed node have no characteristic shape and are typically not symmetric nor unimodal. We consider both the mean and maximum values as summary measures of virulence. Note that the maximum epidemic size is conditional on $N$, but since all nodes are subject to the same number of simulations, we can ignore the conditional in the following discussion. 





\begin{figure}[h!]
\centering
\caption[Comparison of epidemic outcomes between staff and patients]{Comparison of epidemic outcomes between staff and patients. On each box, the horizontal line is the median, the diamond indicates the mean, the edges of the box are the 25th and 75th percentiles, the whiskers extend to the most extreme data points not considered outliers, and outliers are plotted individually. Staff are associated with higher mean epidemic size than patients.} 
\subfloat[Mean epidemic size]{\includegraphics[width =0.5\textwidth]{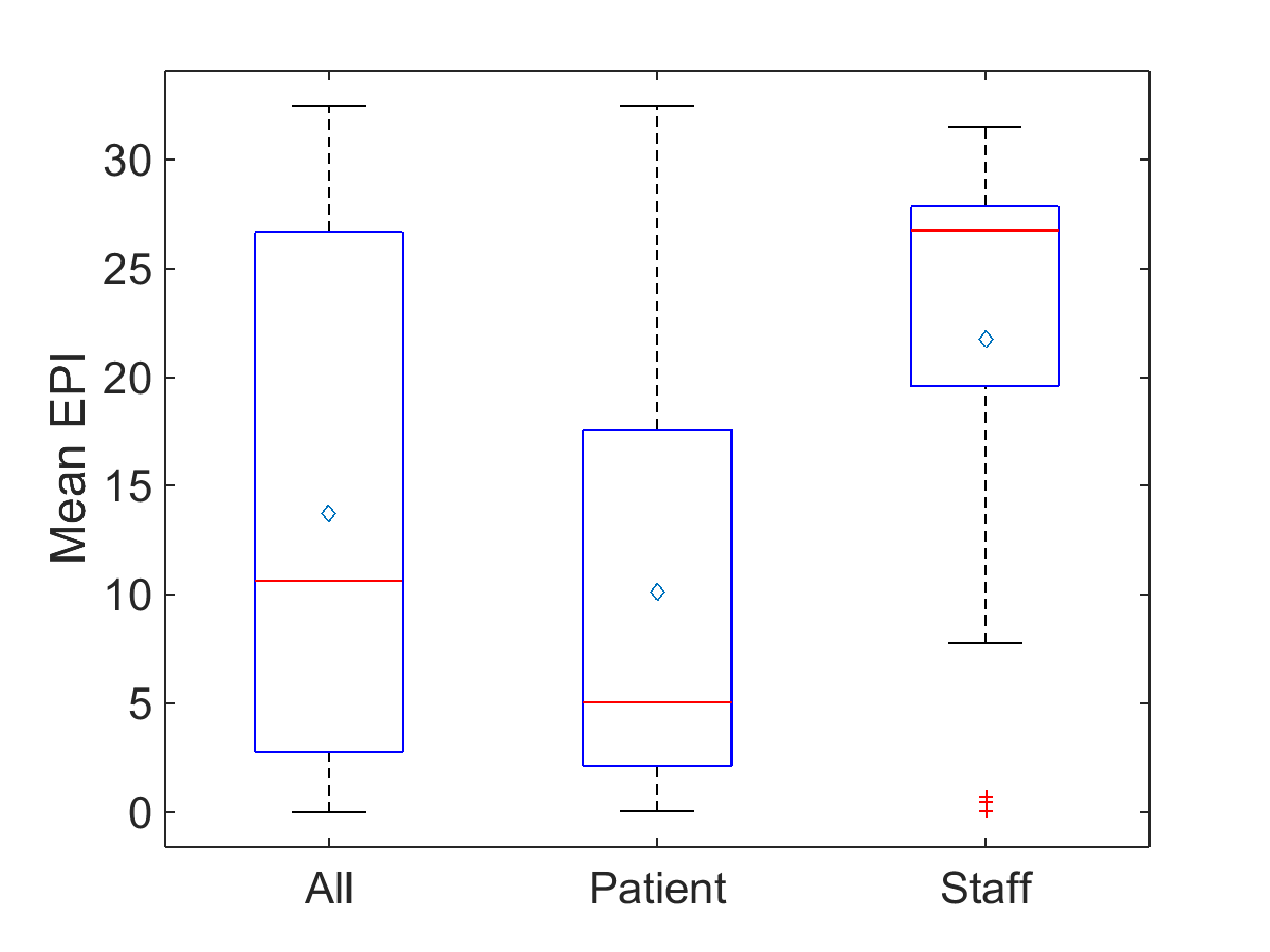}}
\subfloat[Maximum epidemic size]{\includegraphics[width =0.5\textwidth]{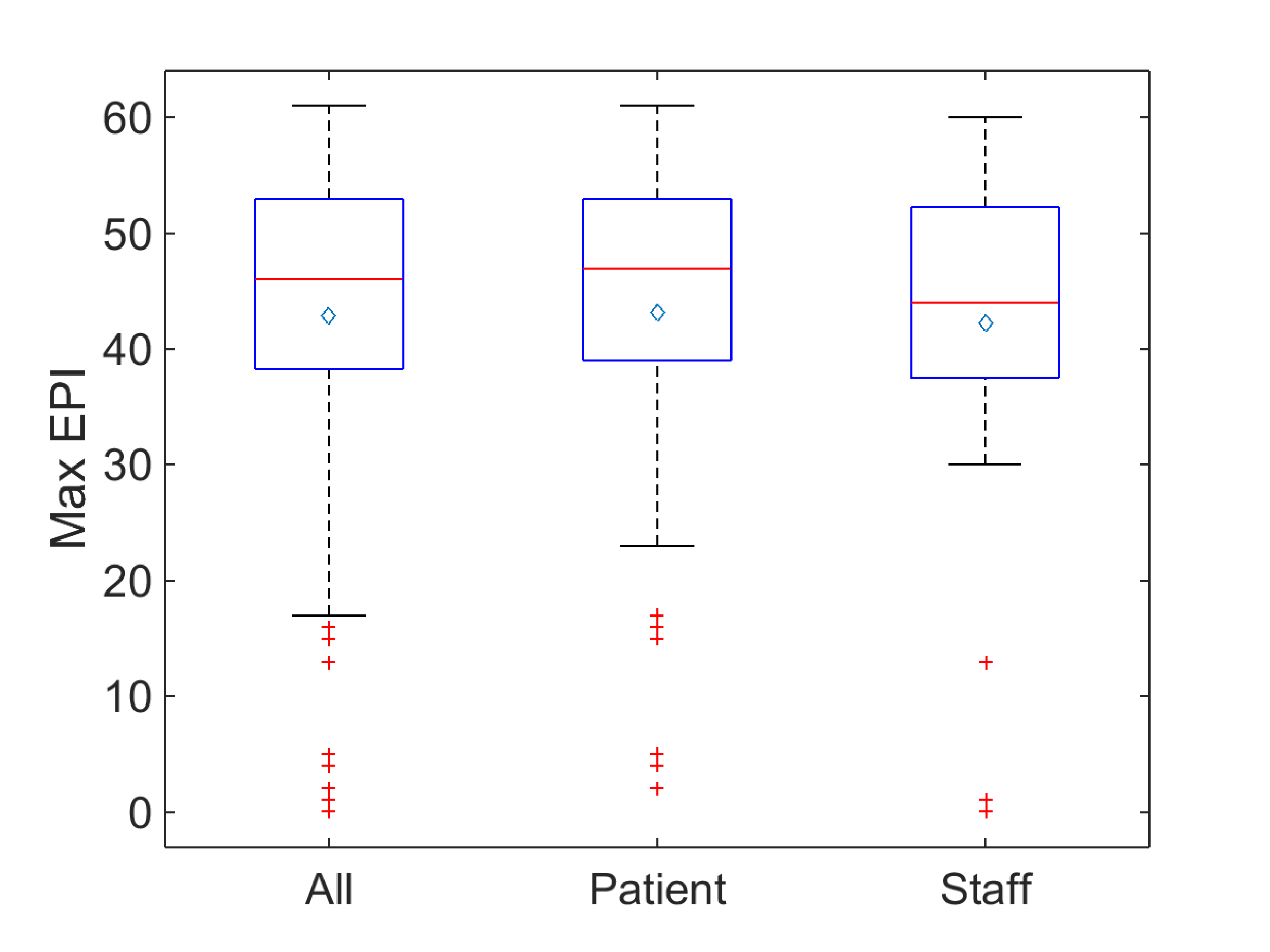}}
\label{boxplots: mean and max epi}
\end{figure}

Figure \ref{boxplots: mean and max epi} displays boxplots of the mean and maximum epidemic size over all initial sources as well as within staff/patient category. Staff members as infection seeds are associated with larger mean epidemic size than patients;  maximum epidemic size is much more similar between the two groups. Worst-case epidemics do not affect more than $60\%$ of the population under study, while on average, less than $31\%$ of the population become infected.  


\section{Relationship between virulence and centrality}
\label{section: virulence+centrality}
In this section we study the relationship between network-based centrality measures and simulated epidemic outcome. We analyze the data  based on shift 1. The centrality measure of interest is the temporal measure based on dynamic communicability. For comparison we consider also the time-aggregated measure AD and the degree based on the binarized aggregated adjacency matrix. The $(i,j)$-th entry of the binarized aggregated adjacency matrix is $1$ if and only if node $i$ and node $j$ made contact at least once during the entire shift; otherwise, the $(i,j)$-th entry is $0$. The degree based on this matrix, so-called binarized degree (BD), therefore ranks nodes based on the number of distinct contacts over time, and therefore does not explicitly contain temporal information. In contrast, aggregated degree AD can be viewed as BD weighted by the duration of contacts, thereby encapsulating more temporal information than BD alone. The discussion in Section \ref{section:Limit} suggests that BC and RC measures based on dynamic communicability can be seen as more nuanced versions of AD, which take into account walks of length $>1$. The question at hand is this: does the increasing complexity of these measures translate into a better understanding of epidemic outcome? We approach this in two ways: the first approach is concerned with identifying top-spreaders, while the second uses linear regression to assess the overall effect of centrality on epidemic outcome. Predictions based on the regression models provide a quantitative means to evaluate the efficacy of centrality measures in the context of epidemic spread.  

Dynamic communicability is computed with $\alpha=0.25*\alpha_{\max}$. As discussed in Section \ref{DC-results}, the BC and RC rankings obtained are fairly robust over the range $\alpha\in [0.25, \, 0.85]* \alpha_{\max}$. We observed that node $20$ (RN) was present in the ED for less than 20 seconds, and made no contact with other nodes during that period. We choose this particular value of $\alpha$ because it was the only one out of four which correctly ranked node $20$ last with respect to both BC and RC measures. 

\subsection{Dynamic communicators}


Nodes are ranked according to epidemic outcome and this is compared with rankings based on network centrality. Table \ref{tables: centrality+virulence} shows that the temporal measure BC outperforms the aggregated measure AD in identifying the most virulent nodes. In comparison to the non-temporal measure BD, BC does a better job in identifying nodes associated with worst-case epidemics (max EPI), but does just as well in identifying nodes associated with large mean epidemic size. None of the high receivers (RC) are among the top 10 virulent nodes, but a subset are among the top 20. 

\begin{figure}[h!]
\centering
\caption[Identification of top spreaders]{Identification of top spreaders. The vertical axis counts the number of nodes ranked highly in terms of both epidemic outcome and centrality (horizontal axis). Including centrality measures in order of complexity increases the coverage of the most virulent nodes.}
\subfloat[Top 10]{\includegraphics[width =0.5\textwidth]{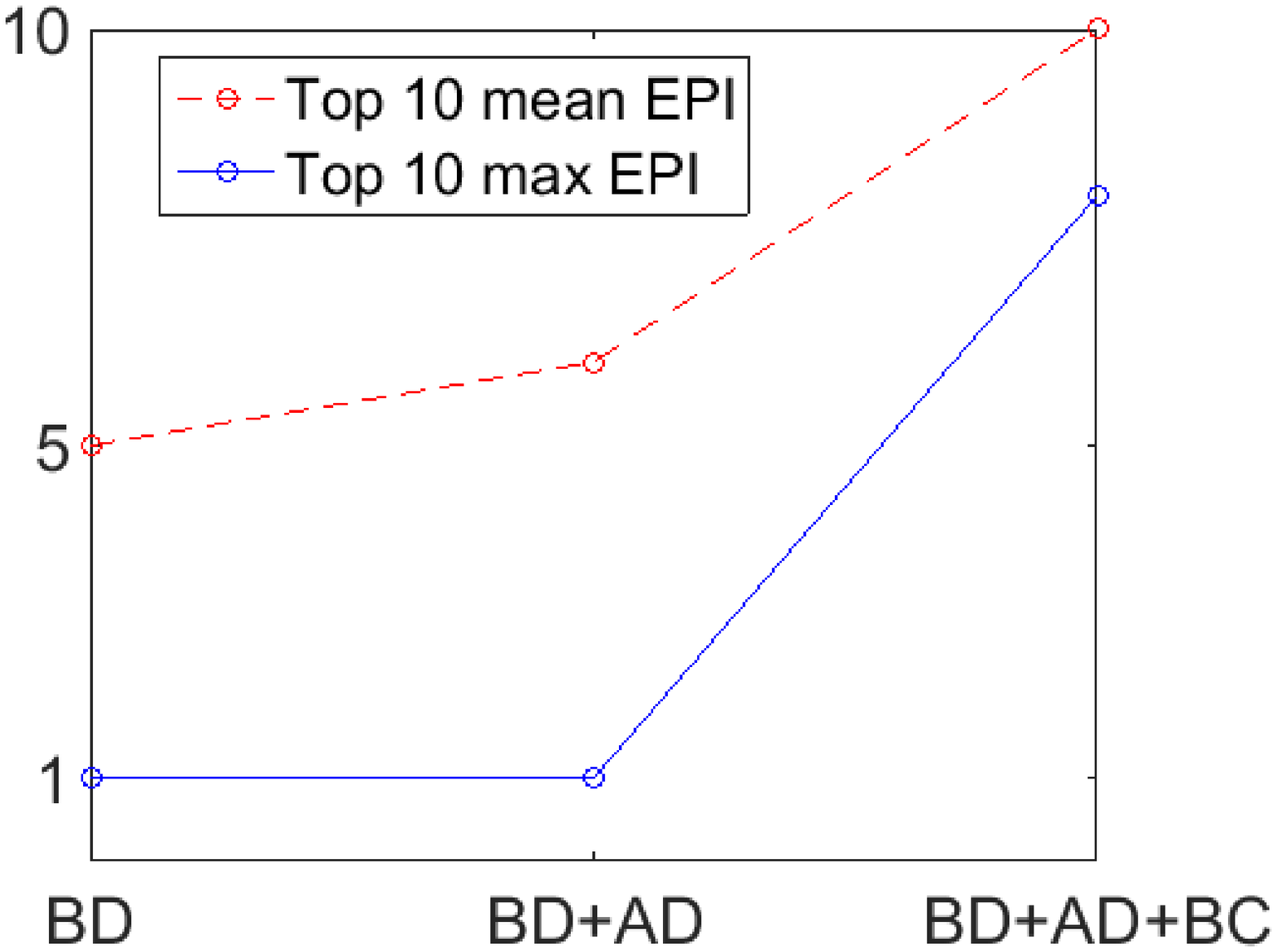}}
\subfloat[Top 20]{\includegraphics[width =0.5\textwidth]{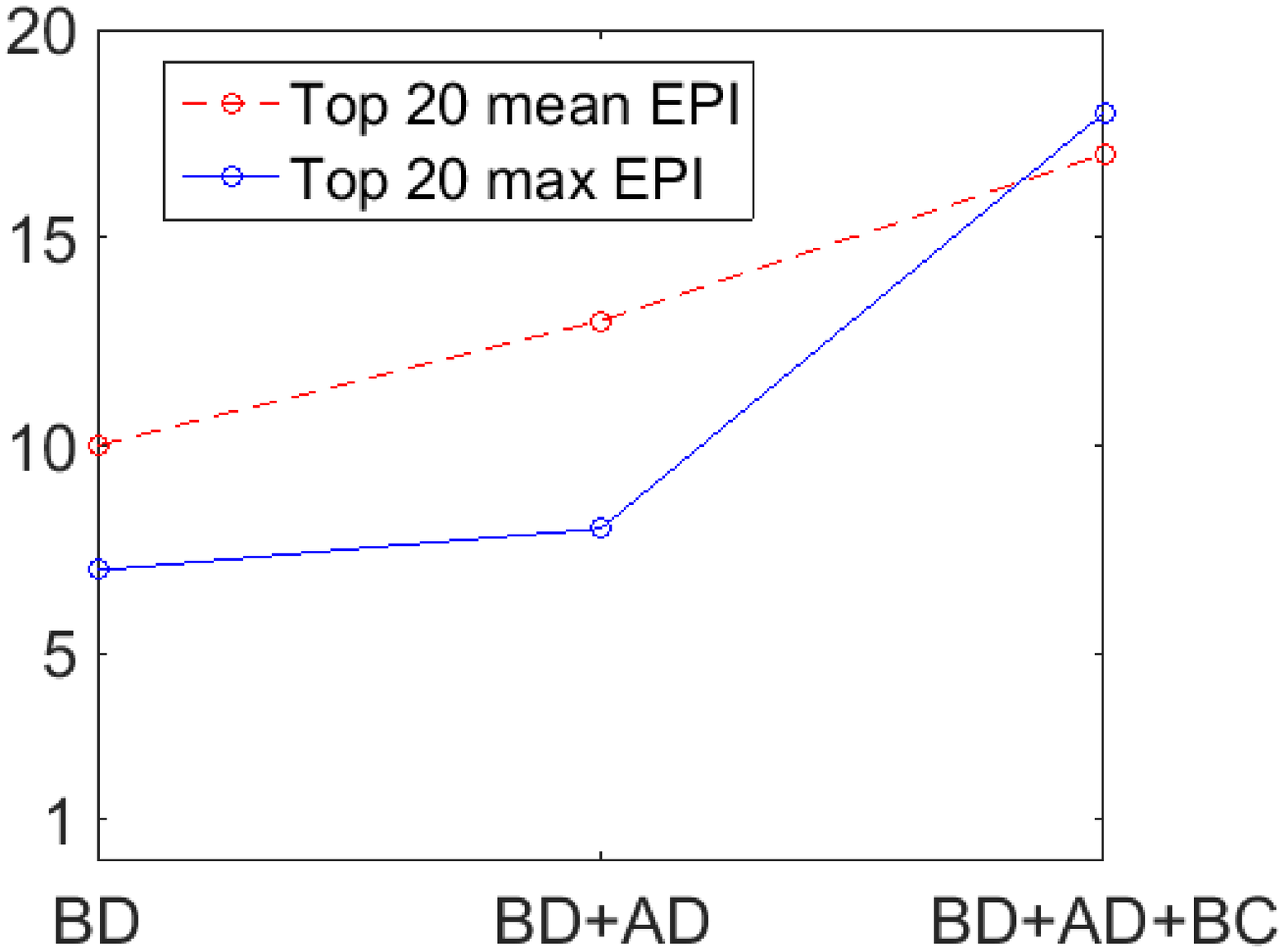}}
\label{figure: top10+20}
\end{figure}

The three different centrality measures identify similar but not identical sets of top-ranked nodes. Dynamic communicators are nodes that are not identified as important by aggregated or non-temporal measures such as AD and BD respectively, but are ranked highly according to temporal measures such as BC (see Section \ref{DC}). Among the top 20 nodes associated with large mean and maximum epidemic sizes, 5 are dynamic communicators in the sense that they rank highly according to BC but not by AD or BD. These 5 nodes are all patients, and among these, 3 have acuity categorized as urgent (the remaining 2 have uncategorized acuity). Utilizing all 3 measures together increases the coverage of top-spreaders as shown in Figure \ref{figure: top10+20}. There is a clear benefit in utilizing BC in addition to BD and AD. Doing so increases the accuracy in capturing virulent nodes from $35\%$ to $90\%$ among the top 10, and from $53\%$ to $88\%$ among the top 20.

\begin{table}[h!]
\footnotesize
\centering
\arraycolsep=1.4pt\def\arraystretch{1.4}
\caption[Comparison of highly central and highly virulent nodes]{Comparison of highly central and highly virulent nodes. Listed are the numbers of nodes which are ranked in the top 10 or 20 by both epidemic outcome and centrality. Note that there may be overlaps in these counts: nodes that are ranked highly by one measure may also be ranked highly with respect to a different measure.}
\label{tables: centrality+virulence}
\begin{tabular}{c|rrrr|rrrr}						\hline\hline												
	&	\multicolumn{4}{c}{Top 10}							&	\multicolumn{4}{c}{Top 20}							\\ \hline	
	&	$BD$	&	$AD$	&	$BC$	&	$RC$	&	$BD$	&	$AD$	&	$BC$	&	$RC$	\\\hline	
mean EPI	&	5	&	1	&	5	&	0	&	10	&	8	&	10	&	7	\\	
max EPI	&	1	&	0	&	8	&	0	&	7	&	2	&	14	&	3	\\\hline \hline	
\end{tabular}

\end{table}

\subsection{Regression analysis}
In this section we study the overall effect of network-based centrality on epidemic outcome. Observe that if a node enters the ED late in the shift, or is present in the ED for a relatively short amount of time, it will have less opportunity to develop connections and consequently, walks across the network, and its network-based centrality measures are likely to suffer as a result. Therefore, possible confounding factors such as the time of first appearance ($T$) and duration observed in the ED ($D$) are included in the analysis, eliminating the need to employ the use of sliding windows in the infection process as is common in the literature \cite{Higham_predicts, Rocha-SexNetwork}. Staff/patient category is also included as a predictor in the model. 

We point out that we also computed BC and RC measures based on this version of dynamic communicability: 
\[\hat{Q} = (I+\alpha A^{[1]})(I+\alpha A^{[2]})\cdots (I+\alpha A^{[M]}).\]
Recall that this imposes the restriction that there is at most one edge per time-step. This modification only slightly improved the fit of the data (improvements, if any, were in the fourth decimal place), suggesting that in this application, imposing such a restriction makes little difference to BC and RC measures and in particular, has little impact on explaining overall epidemic outcome.

We regress the epidemic outcome (mean/maximum EPI) on the centrality measures (BC, log(BC), AD, BD) while adjusting for the time of first appearance in the ED ($T$), duration active in the ED ($D$) and the staff binary indicator ($S$). We perform standard linear regression (using Matlab's \texttt{fitlm}), where the error terms associated with the responses are assumed to be independent and normally distributed with mean equal to 0. To have the covariates on a similar scale, we standardize the AD and BD values. All other quantitative covariates are mean-centered to assist in the interpretation of the coefficients. Model coefficients and associated $95\%$ confidence intervals are reported in Table \ref{tables: model-summary}.  

\begin{table}[h!]
\footnotesize
\centering
\arraycolsep=1.4pt\def\arraystretch{1.4}
\caption[Regression analysis summary]{Estimated regression coefficients and associated $95\%$ confidence intervals for the eight models under consideration. The response/dependent variable is one of mean EPI or max EPI associated with the seed node; the predictor of interest is the seed node's centrality measure. Measures studied are BC, log(BC), AD and BD. AD and BD values are standardized; all  other quantitative covariates are mean-centered.}
\begin{adjustbox}{max width=\textwidth}
\begin{tabular}{l|cccc|cccc} \hline\hline																	
	&	\multicolumn{4}{c}{mean EPI}							&	\multicolumn{4}{c}{max EPI}							\\ \hline
intercept	&	$10.90^*$	&	$10.92^*$	&	$12.24^*$	&	$11.92^*$	&	$43.17^*$	&	$43.24^*$	&	$43.36^*$	&	$43.68^*$	\\
	&	(8.97, 12.83)	&	(8.84, 12.99)	&	(10.10, 14.37)	&	(10.15, 13.68)	&	(41.91, 44.44)	&	(41.99, 44.49)	&	(41.96, 44.77)	&	(42.49, 44.87)	\\
$BC$	&	$40.17^*$	&		&		&		&	$16.29^*$	&		&		&		\\
	&	(22.09, 58.25)	&		&		&		&	(4.42, 28.16)	&		&		&		\\
$\log(BC)$	&		&	0.04	&		&		&		&	$0.04^*$	&		&		\\
	&		&	(-0.01, 0.08)	&		&		&		&	(0.02, 0.07)	&		&		\\
$AD$ (std)	&		&		&	$3.79^*$	&		&		&		&	0.59	&		\\
	&		&		&	(1.67, 5.92)	&		&		&		&	(-0.81, 1.98)	&		\\
$BD$ (std)	&		&		&		&	$6.53^*$	&		&		&		&	$3.22^*$	\\
	&		&		&		&	(4.65, 8.41)	&		&		&		&	(1.95, 4.49)	\\
$D$ (in hours)	&	$1.02^*$	&	$0.66^*$	&	0.56	&	-0.17	&	0.34	&	0.11	&	0.21	&	-0.22	\\
	&	(0.43, 1.62)	&	(0.02, 1.31)	&	(-0.06, 1.17)	&	(-0.76, 0.43)	&	(-0.05, 0.73)	&	(-0.27, 0.50)	&	(-0.19, 0.62)	&	(-0.62, 0.18)	\\
$T$ (in hours)	&	$-0.83^*$	&	0.68	&	-1.29	&	$-1.04^*$	&	$-3.94^*$	&	$-1.88^*$	&	$-4.13^*$	&	$-4.00^*$	\\
	&	(-1.43, -0.23)	&	(-1.70, 3.06)	&	(-1.86, -0.71)	&	(-1.55, -0.54)	&	(-4.33, -3.54)	&	(-3.32, -0.45)	&	(-4.50, -3.75)	&	(-4.34, -3.66)	\\
$S$	&	$9.09^*$	&	$9.05^*$	&	$4.77^*$	&	$5.81^*$	&	-1.05	&	-1.26	&	-1.66	&	$-2.69^*$	\\
	&	(5.25, 12.94)	&	(4.91, 13.20)	&	(0.06, 9.48)	&	(2.20, 9.42)	&	(-3.58, 1.47)	&	(-3.76, 1.24)	&	(-4.76, 1.43)	&	(-5.13, -0.26)	\\\hline
$R^2$	&	0.54 (0.12)	&	0.46 (0.23)	&	0.51 (0.31)	&	0.62 (0.54)	&	0.87 (0.14)	&	0.87 (0.86)	&	0.86 (0.01)	&	0.89 (0.17)	\\\hline\hline
\multicolumn{1}{}{} $^* p<0.05$																	\\
\multicolumn{5}{}{} $R^2$ values in parantheses indicate the values obtained in single-predictor models																	\\ \hline
\end{tabular}																	

\end{adjustbox}
\label{tables: model-summary}
\end{table}

We report $R^2$ as a measure of goodness-of-fit. $0\leq R^2 \leq 1$ can be thought of as a generalized form of squared Pearson correlation when there is more than one predictor. A high value of $R^2$ is indicative of a good fit to the data. With the exception of max EPI $\sim \log(BC)$, low $R^2$ values associated with single-predictor models (see $R^2$ values in parentheses in Table \ref{tables: model-summary}) suggest that centrality measures alone cannot fully explain epidemic outcome, while the inclusion of $D$ and $T$ improve the fit to the data.

Coefficients of the centrality measures are $>0$, suggesting that an increase in centrality score is associated with higher epidemic outcome. As expected, there is typically a positive estimated effect of $D$ on epidemic outcome: the longer the seed node is active in the ED, the larger the epidemic outcome. On the other hand, there is typically a negative estimated effect associated with $T$: the later the seed node arrives in the ED, the smaller the associated epidemic outcome\footnote{The positive coefficient for $T$ when regressing mean EPI on $\log(BC)$ may be due to interaction effects --- see Table \ref{tables: interaction by A}.}. The intercept term is interpreted as the expected epidemic outcome among patients with average values of centrality, $T$ and $D$. Among staff members with average centrality, $T$ and $D$, the expected mean epidemic size is significantly larger (by 9.09, 9.05, 4.77, 5.81) than that associated with patients; with respect to maximum epidemic size, the reverse is true: staff members are associated with slightly smaller maximum epidemic sizes compared to patients, but these differences are not significant.    

For each of the models in Table \ref{tables: model-summary}, we report the interaction effects of the confounders separately. In Table \ref{tables: interaction by D} we see that the coefficients are typically positive, suggesting that regardless of $D$, an increase in centrality is associated with an increase in epidemic outcome. Strong monotonically decreasing interaction effects are observed for $AD$: the longer nodes are active in the ED (larger $D$), the smaller the effect of $AD$ on epidemic outcome. On the other hand, the effect of $AD$ on epidemic outcome increases with $T$ (Table \ref{tables: interaction by A}). This suggests that minimizing $AD$ among later arrivals may reduce mean epidemic size. The reverse phenomenon is observed for log(BC): the later a node appears in the ED, the smaller the effect of log(BC) on epidemic outcome. It is interesting to note that among late arrivals, a unit increase in BD (or equivalently, an additional distinct contact) on average results in a relatively large increase in maximum epidemic size. Nodes in the higher quartiles of $T$ having the same BC, $D$ and $S$ values resulted in a rank deficient design matrix --- coefficient estimates are not reported in this case. In Table \ref{tables: interaction by staff} we see that the effect of centrality (apart from $BC$) is similar among both staff and patients.   

\begin{table}[h!]
\footnotesize
\centering
\arraycolsep=1.4pt\def\arraystretch{1.4}
\caption[Interaction by $D$ (in hours)]{Interaction by $D$ (in hours) is examined by 
regressing epidemic outcome with respect to the centrality measure, stratified by duration ($D$) in hours, while adjusting for $T$ and $S$. The coefficients associated with the centrality measure for each group are reported below.}
\begin{tabular}{c|c|cccc} \hline\hline											
	&		&	$0<D\leq 2.51$	&	$2.51<D\leq 4.78$	&	$4.78<D\leq 8.05$	&	$D>8.05$	\\\hline
\multirow{8}{*}{mean EPI}	&	$BC$	&	$52.55^*$	&	$56.15^*$	&	21.77	&	4.68	\\
	&		&	(39.37, 65.73)	&	(1.63, 110.66)	&	(-22.91, 66.44)	&	(-51.60, 60.97)	\\
	&	$\log(BC)$	&	0.01	&	0.08	&	0.05	&	0.27	\\
	&		&	(-0.08, 0.10)	&	(-0.01, 0.18)	&	(-0.10, 0.19)	&	(-0.11, 0.65)	\\
	&	$AD$	&	$125.67^*$	&	$36.37^*$	&	5.10	&	2.23	\\
	&		&	(97.45, 153.90)	&	(16.43, 56.31)	&	(-3.19, 13.40)	&	(-0.16, 4.61)	\\
	&	$BD$	&	$7.03^*$	&	$9.06^*$	&	$10.19^*$	&	$3.35^*$	\\
	&		&	(2.15, 11.91)	&	(3.64, 14.47)	&	(6.41, 13.96)	&	(0.41, 6.30)	\\ \hline
\multirow{8}{*}{max EPI}	&	$BC$	&	9.76	&	42.26	&	7.38	&	$26.97^*$	\\
	&		&	(-9.92, 29.43)	&	(-0.48, 85.01)	&	(-11.77, 26.53)	&	(1.97, 51.98)	\\
	&	$\log(BC)$	&	$0.07^*$	&	0.04	&	0.04	&	-0.11	\\
	&		&	(0.01, 0.13)	&	(-0.04, 0.11)	&	(-0.02, 0.10)	&	(-0.30, 0.07)	\\
	&	$AD$	&	$47.89^*$	&	16.62	&	0.51	&	-0.48	\\
	&		&	(5.78, 90.00)	&	(-1.79, 35.04)	&	(-3.14, 4.15)	&	(-1.72, 0.76)	\\
	&	$BD$	&	$6.73^*$	&	$7.69^*$	&	$3.70^*$	&	0.25	\\
	&		&	(3.42, 10.05)	&	(3.69, 11.70)	&	(1.82, 5.59)	&	(-1.36, 1.85)	\\ \hline\hline
\multicolumn{1}{}{} $^* p<0.05$											\\\hline
\end{tabular}											

\label{tables: interaction by D}
\end{table}

\begin{table}[h!]
\footnotesize
\centering
\arraycolsep=1.4pt\def\arraystretch{1.4}
\caption[Interaction by $T$ (in hours)]{Interaction by $T$ (in hours) is examined by 
regressing epidemic outcome with respect to the centrality measure, stratified by $T$ in hours, while adjusting for $D$ and $S$. The coefficients associated with the predictor of interest for each group are reported below.}
\begin{tabular}{c|c|cccc} \hline\hline											
	&		&	$0<T\leq 1.14$	&	$1.14<T\leq 2.25$	&	$2.25<T\leq 5.65$	&	$T>5.65$	\\\hline
\multirow{8}{*}{mean EPI}	&	$BC$	&	$55.22^*$	&	30.97	&	---	&	--- 	\\
	&		&	(35.97, 74.46)	&	(-320.73, 382.66)	&	---	&	---	\\
	&	$\log(BC)$	&	$2.36^*$	&	0.28	&	0.03	&	0.01	\\
	&		&	(0.92, 3.81)	&	(-0.16, 0.71)	&	(-0.07, 0.12)	&	(-0.01, 0.02)	\\
	&	$AD$	&	1.60	&	2.71	&	3.91	&	$6.00^*$	\\
	&		&	(-4.68, 7.88)	&	(-2.20, 7.62)	&	(-0.05, 7.87)	&	(3.37, 8.64)	\\
	&	$BD$	&	$9.54^*$	&	$4.58^*$	&	$7.98^*$	&	$3.71^*$	\\
	&		&	(4.68, 14.40)	&	(0.86, 8.30)	&	(3.81, 12.15)	&	(0.77, 6.65)	\\\hline\hline
\multirow{8}{*}{max EPI}	&	$BC$	&	$22.36^*$	&	11.93	&	---	&	--- 	\\
	&		&	(8.28, 36.44)	&	(-171.06, 194.91)	&	--- 	&	--- 	\\
	&	$\log(BC)$	&	$1.89^*$	&	$0.38^*$	&	0.04	&	$0.06^*$	\\
	&		&	(1.34, 2.45)	&	(0.21, 0.55)	&	(-0.00, 0.08)	&	(0.04, 0.09)	\\
	&	$AD$	&	-0.90	&	-0.60	&	-0.66	&	0.89	\\
	&		&	(-4.40, 2.60)	&	(-3.21, 2.01)	&	(-2.65, 1.33)	&	(-7.35, 9.13)	\\
	&	$BD$	&	$3.84^*$	&	1.65	&	$3.06^*$	&	$13.33^*$	\\
	&		&	(0.70, 6.98)	&	(-0.42, 3.72)	&	(0.91, 5.21)	&	(8.56, 18.11)	\\ \hline\hline
\multicolumn{1}{}{} $^* p<0.05$											\\
\multicolumn{3}{}{} --- design matrix is rank deficient 											\\ \hline
\end{tabular}

\label{tables: interaction by A}
\end{table}

\begin{table}[h!]
\footnotesize
\centering
\arraycolsep=1.4pt\def\arraystretch{1.4}
\caption[Interaction by staff/patient category]{Data is stratified according to staff/patient category. Within each group, we regress epidemic outcome with respect to the centrality measure, while adjusting for $D$ and $T$. The coefficients associated with the centrality measure for each group are reported below.}
\begin{tabular}{c|c|cc} \hline\hline							
	&		&	staff	&	patient	\\\hline
\multirow{8}{*}{mean EPI}	&	$BC$	&	3.21	&	$45.21^*$	\\
	&		&	(-36.07, 42.49)	&	(24.15, 66.28)	\\
	&	$\log(BC)$	&	$0.05^*$	&	0.04	\\
	&		&	(0.01, 0.09)	&	(-0.02, 0.10)	\\
	&	$AD$	&	$2.43^*$	&	$36.10^*$	\\
	&		&	(1.07, 3.79)	&	(26.06, 46.14)	\\
	&	$BD$	&	1.02	&	$9.86^*$	\\
	&		&	(-1.80, 3.83)	&	(7.46, 12.25)	\\\hline
\multirow{8}{*}{max EPI}	&	$BC$	&	15.13	&	$16.55^*$	\\
	&		&	(-13.24, 43.50)	&	(2.66, 30.44)	\\
	&	$\log(BC)$	&	0.03	&	$0.05^*$	\\
	&		&	(-0.00, 0.06)	&	(0.02, 0.09)	\\
	&	$AD$	&	0.13	&	$11.67^*$	\\
	&		&	(-1.08, 1.34)	&	(4.10, 19.24)	\\
	&	$BD$	&	0.63	&	$5.03^*$	\\
	&		&	(-1.45, 2.71)	&	(3.37, 6.69)	\\\hline\hline
\multicolumn{1}{}{} $^* p<0.05$							\\ \hline
\end{tabular}							

\label{tables: interaction by staff}
\end{table}

\label{Regression}

\subsection{Prediction}
\label{Prediction}
The regression models from Section \ref{Regression} were used to predict simulations from shifts 2 to 7. Simulated outcomes are considered to be the ground-truth. 
Predictions are performed separately for each shift. Predicted values ($\hat{Y}$) are compared to the values obtained by simulation ($Y$). Prediction errors are computed in two ways (where $n$ is the total number of nodes):
\begin{align*}
\text{RMSE} &= \sqrt{\frac{1}{n}\sum_{i=1}^n \left(\hat{Y}_i - Y_i\right)^2} \\
\text{BIAS} &= \frac{1}{n}\sum_{i=1}^n \left(\hat{Y}_i - Y_i\right)
\end{align*} 
Since the range of the epidemic outcomes differ by shift, we standardize RMSE and BIAS values by multiplication by $100/n$. The average of the standardized RMSE and BIAS values over 6 shifts are reported in Table \ref{table: RMSE+BIAS_full}. A standardized RMSE value of 10.32 means that, on average, the predicted epidemic outcome (mean/max EPI) fails to correctly capture the observed outcome by a factor of $10.32\%$ of the total population under study. The sign of the bias provides an indication of over/under-estimation: a positive bias of 1.45 means that on average, the predicted outcome overestimates the observed outcome by $1.45\%$ of the population. From the perspective of predicting epidemic outcome, a positive bias is preferable to a negative bias, since one would prefer to err on the side of caution. From Table \ref{table: RMSE+BIAS_full}, we can see that for all the models under consideration, predicted values overestimate the observed outcome. In addition, within this framework, predictions of the mean epidemic size are consistently more accurate than predictions of the maximum epidemic size. 

To assess the impact of including network centralities in developing predictive models, we consider models without centrality: that is, we regress the response against $D$, $T$ and $S$ only. The difference in prediction error relative to the null model quantifies the change in predictive power: a decrease in RMSE/BIAS values indicate a stronger predictive model compared to the null model. The percentage changes in prediction error relative to the null model are reported in parentheses in Table \ref{table: RMSE+BIAS_full}. Inclusion of centrality reduces RMSE by $0.04 - 10\%$, while BIAS increases, albeit only slightly. This means that the inclusion of centrality results in predictions that tend to slightly increase overestimation of the true values, but the predictions are overall more accurate. 

\begin{figure}[h!]
\centering
\caption[Predictions based on full models (shift 5)]{Predictions based on full models (shift 5). Observed values are denoted `o', predicted values are denoted `+' (color online). }
\label{fig: prediction_full}
\includegraphics[scale=0.5]{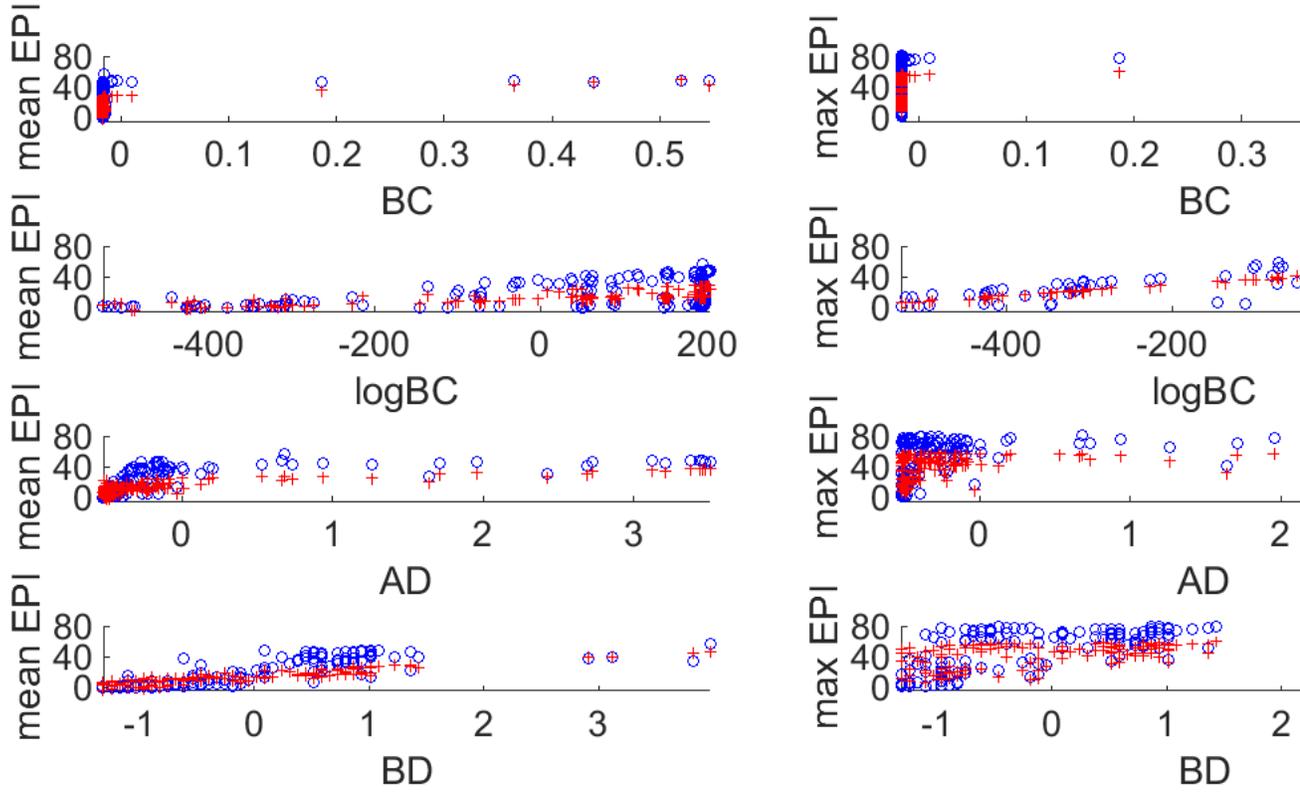}
\end{figure}

Figure \ref{fig: prediction_full} displays the predictions (based on the full models) relative to the simulated outcomes for shift 5. The full model generates predictions that capture the overall trend of the data very well, and there is significant improvement compared to the predicted trends based on single-predictor models (not shown). 
Reduction in RMSE suggests that inclusion of network-based centrality does improve predictive power, but the improvement is marginal, and whether or not this level of improvement is worth the computational effort remains subjective. Regardless, it is clear that temporal markers such as $D$ (duration observed) and $T$ (time of first appearance) play an important role in reproducing realistic prediction patterns. 


A key observation is that in terms of predictive power, inclusion of the temporal measure BC or log(BC) does not necessarily outperform the aggregated/non-temporal measures AD and BD. With respect to both mean and maximum epidemic size, BD is associated with the largest decrease in RMSE compared to the null model. This may be due to the fact that BD has the strongest linear relationship with epidemic outcome (see $R^2$ values in parentheses in Table \ref{tables: model-summary}) compared to the other centrality measures. With respect to maximum epidemic size, inclusion of log(BC) improves the quality of the predictions more than the inclusion of AD, but does not outperform the non-temporal measure BD.  

\begin{table}[h!]
\centering
\footnotesize
\arraycolsep=1.4pt\def\arraystretch{1.4}
\caption[Prediction errors for full models]{Prediction errors for full models. Reported values are standardized and averaged over 6 shifts. Values in parentheses are the percentage changes compared to the prediction errors based on the null model (where the response is regressed against $D$, $T$ and $S$ only).}
\label{table: RMSE+BIAS_full}
\begin{tabular}{c| llll | l | l} \hline\hline													
Response	&	\multicolumn{4}{c|}{Predictors}							&	RMSE	&	BIAS	\\\hline
\multirow{5}{*}{mean EPI}	&	$BC$	&	D	&	T	&	S	&	10.32 ($-2.02\%$)	&	1.45 ($0.27\%$)	\\
	&	$\log(BC)$	&	D	&	T	&	S	&	10.34 ($-1.78\%$)	&	1.45 ($0.31\%$)	\\
	&	$AD$	&	D	&	T	&	S	&	9.99 ($-5.14\%$)	&	1.52 ($5.40\%$)	\\
	&	$BD$	&	D	&	T	& 	S	&	9.44 ($-10.33\%$)	&	1.50 ($4.17\%$)	\\
	&		&	D	&	T	&	S	&	10.53	&	1.44	\\ \hline 
\multirow{5}{*}{max EPI}	&	$BC$	&	D	&	T	&	S	&	15.57 ($-0.04\%$)	&	4.26 ($0.04\%$)	\\
	&	$\log(BC)$	&	D	&	T	&	S	&	15.36 ($-1.38\%$)	&	4.27 ($0.12\%$)	\\
	&	$AD$	&	D	&	T	&	S	&	15.55 ($-0.16\%$)	&	4.27 ($0.28\%$)	\\
	&	$BD$	&	D	&	T	&	S	&	15.10 ($-3.10\%$)	&	4.29  ($0.70\%$)	\\
	&		&	D	&	T	&	S	&	15.58	&	4.26	\\ \hline \hline
\end{tabular}

\end{table}


\section{Conclusions and future work}
\label{Conclusion}
We have conducted a thorough case-study of dynamic communicability on an empirical temporal contact network. Computation of the dynamic communicability matrix $Q$ requires a choice of parameter $\alpha$. We have demonstrated the robustness of the node rankings obtained with respect to the choice of $\alpha$ in the range $[0.25,\,0.85]*\alpha_{\max}$. To avoid overflow, normalization was performed at each time-step, resulting in centrality measures that are numerically tiny. In spite of this, convergence to the aggregated measure AD is also observed empirically, so even though the computed centrality measures themselves are small, we are fairly confident that the associated rankings are correct. Furthermore, the rankings obtained for $\alpha$ in the range $[0.25,\,0.85]*\alpha_{\max}$ are significantly different from the rankings based on AD, suggesting that within this range of $\alpha$ the temporal measures BC and RC provide new insights that cannot be attained from the time-aggregated point of view.    

Analysis on the data revealed that high receivers are typically staff. A subset of RN's have particularly high receiving scores relative to the other nodes in the network. On the other hand, patients are on the whole slightly stronger broadcasters than staff. Staff (as seed nodes of infection) are associated with larger mean epidemic size than patients, but worst-case epidemics are similarly distributed across both groups. 
 
Analysis shows that the non-temporal measure BD identifies a largely different set of top-spreaders from the temporal measure BC. Taken together, both measures are able to correctly identify a large proportion of highly virulent nodes. The aggregated measure AD also adds value, but less significantly. We propose using all 3 measures to increase overall accuracy in identifying nodes associated with large epidemic outcome. Such an approach may increase the efficacy of disease-prevention/mitigation strategies.    


Our results show that network-based centralities identify a good proportion of top-spreaders, but fail to identify all of them: there are low-ranking nodes associated with large epidemic outcomes. This is in agreement with the work of \cite{influence_of_all_nodes, centrality_underestimates} and \cite{CompleNet}. Network centrality can therefore help elucidate properties of nodes (such as their broadcasting ability) but centrality alone cannot fully account for the observed patterns of epidemic spread. Specifically, with respect to the infection model employed in this study, network centrality on its own cannot be used as a proxy for time-intensive simulations as claimed in \cite{Higham_predicts}. 

Instead of using raw correlations to quantify the relationship between centrality and epidemic spread, we use linear regression to study the overall effect of centrality on epidemic outcome, while adjusting for possible confounding factors. We find that all forms of centrality studied (BC, AD, BD) have an estimated positive effect on epidemic outcome. There is also evidence of interaction effects between centrality and other variables such as time of first appearance ($T$) and duration observed in the ED ($D$). Inclusion of centrality in the model reduces prediction error, and in this regard,  BD outperforms the other temporal measures. The aggregated measure AD also arguably does better than the more nuanced measure BC: our results show that the additional information captured by temporal centrality does not appear to have a substantial effect within this linear regression framework. The ability of BC to identify top-spreaders is smoothed out in the averaging process of linear regression. RMSE and BIAS values used to quantify predictive power are also averaged across all individuals. Nonetheless, the predictive models capture the trend of the data remarkably well. Improvement relative to the null model (without centrality) is marginal, and whether or not this level of improvement is worth the computational effort remains subjective. Regardless, our results also indicate that other temporal markers such as duration observed ($D$) and the time of first appearance ($T$) are essential in reproducing realistic patterns of epidemic outcome.  


Our approach of using network centrality as a predictor for epidemic outcome in a linear regression model 
is a first attempt, and we believe that more sophisticated modeling techniques may be able to take advantage of the nuances provided by temporal centrality to improve predictive power. A practical difficulty encountered is the fact that the temporal measures BC and RC fail to differentiate a large majority of the nodes. A measure incorporating temporal information that is at the same time less highly-skewed may improve predictive performance. The normalization proposed in \cite{e.colman} may help to normalize the distribution of centrality scores. We also point out that the techniques in \cite{e.colman} present another way to deal with the confounding temporal effects on BC and RC scores which are worth exploring in future work.  


Another issue is that the summary measures of mean and maximum epidemic size may be too crude an estimate of virulence. Other measures such as the total number of infection paths, or the time taken to reach a certain epidemic threshold can also be incorporated into the analysis. 

The choice of time-scale at which to aggregate the contact matrices remains to be studied in greater detail. Preliminary analysis shows that aggregation into 20 minute time frames results in better distinction among top RC nodes. This is in some sense surprising, because at face-value, aggregation would seem to make nodes `more equal': within a 20 minute interval, there is no distinction between a contact pair lasting for one second and a contact pair lasting for the full 20 minutes. However, aggregation also results in fewer contact matrices, and therefore less severe normalization effects when computing $Q$. This competing effect may explain why aggregation into 20 minute time frames results in a more nuanced differentiation among top RC nodes. All three temporal measures, BC, RC and AD, are expected to be sensitive to the time-scale chosen, and a balance needs to be drawn between the loss of temporal information on the one hand, and the computational benefits of having fewer contact matrices to work with on the other hand.

As we analyze the results of our findings, it is important to bear in mind the limitations 
of our study. People monitored were allowed to move freely in and out of the ED, but contact data does not include potential disease-transmitting interactions that took place beyond the designated zones. There were also staff and patients in the ED who did not participate in the study. Therefore the infection simulations necessarily underestimate the true epidemic outcomes. Our study also does not take into account other routes of disease transmission, such as indirect fomite transmission via shared objects (such as door handles). Including locations in the disease-modeling aspect as well as in the network-based centrality can potentially provide different insights.





\section*{Acknowledgements}
We thank George Cotsonis for providing the data, Lisa Elon for helpful discussions at early stages of the project, and Andres Celis for writing the code for the infection simulation. We also thank an anonymous reviewer for helpful comments.  

\bibliographystyle{plain}
\bibliography{./wns-bib}

\end{document}